\documentclass[10pt,a4paper]{article}
\usepackage[left=2cm,right=2cm,top=2.5cm,bottom=2.5cm]{geometry}
\usepackage{graphicx}

\usepackage[T1]{fontenc}
\usepackage{times}  

\usepackage{amsmath}
\usepackage{bm}
\usepackage[numbers,sort&compress]{natbib}
\usepackage{float}
\usepackage[colorlinks=true,
  linkcolor=black, citecolor=blue, urlcolor=blue]{hyperref}   
\usepackage[dvipsnames,svgnames,x11names,hyperref]{xcolor}  
\usepackage[capitalize]{cleveref}

\usepackage{setspace}
\crefformat{subsection}{\S#1}

\title{\textbf{Towards fully predictive gyrokinetic full-f simulations: validation and triangularity studies in TCV}}

\author{
  A. C. D. Hoffmann$^{1}$,
  T. N. Bernard$^{2}$,
  M. Francisquez$^{1}$,
  G. W. Hammett$^{1}$,
  A. Hakim$^{1}$,\\
  J. Boedo$^{3}$,
  R. Rizkallah$^{3}$,
  C. K. Tsui$^{4}$,
  and the TCV team$^{5}$
  \\[0.8em]
  \normalsize
  $^{1}$ Princeton Plasma Physics Laboratory, Princeton, NJ, USA\\
  $^{2}$ General Atomics, San Diego, CA, USA\\
  $^{3}$ University of California San Diego, La Jolla, CA, USA\\
  $^{4}$ Sandia National Laboratories, Albuquerque, NM, USA\\
  $^{5}$ See author list of B.P. Duval et al 2024 Nucl. Fusion 64 112023
  \\[0.5em]
  E-mail: ahoffmann@pppl.gov
}
\date{}

 \newcommand{\vpar}{v_\parallel}

 \newcommand{\grad}{\nabla}

\newcommand{\ExB}{\bm E\times\bm B}

\newcommand{\gyacomo}{\textsc{Gyacomo}}
\newcommand{\gkyl}{\textsc{Gkeyll}}

\newcommand{\modifi}[1]{{#1}}

\usepackage{natbib}
\usepackage{graphicx}
\usepackage{bm}
\usepackage{cancel}
\usepackage{float}
\usepackage{algorithm}
\usepackage{algpseudocode}
\usepackage{tikz}
\usepackage{txfonts}

\begin{document}

\maketitle

\begin{abstract}
Designing economical magnetic confinement fusion power plants motivates computational tools that can estimate plasma behavior from engineering parameters without direct reliance on experimental measurement of the plasma profiles. 
In this work, we present full-$f$ global \modifi{long-wavelength} gyrokinetic simulations of edge and scrape-off layer turbulence in tokamaks that use only magnetic geometry, heating power, and particle inventory as inputs. 
Unlike many modeling approaches that employ free parameters fitted to experimental data, raising uncertainties when extrapolating to reactor scales. This approach directly simulates turbulence and resulting profiles through gyrokinetics without such empirical adjustments.
This is achieved via an adaptive sourcing algorithm in \gkyl{} that strictly controls energy injection and emulates particle sourcing due to neutral recycling. 
We show that the simulated kinetic profiles compare reasonably well with Thomson scattering and Langmuir probe data for Tokamak à Configuration Variable (TCV) discharge \#65125, and that the simulations reproduce characteristic features such as blob transport and self-organized electric fields.
Applying the same framework to study triangularity effects suggests mechanisms contributing to the improved confinement reported for negative triangularity (NT). Simulations of TCV discharges \#65125 and \#65130 indicate that NT increases the $\mathbf{E} \times \mathbf{B}$ flow shear (by about 20\% in these cases), which correlates with reduced turbulent losses and a modest change in the distribution of power exhaust to the vessel wall.
While the physical models contain approximations that can be refined in future work, the predictive capability demonstrated here, evolving multiple profile relaxation times with kinetic electron and ion models in hundreds of GPU hours, indicates the feasibility of using \gkyl{} to support design studies of fusion devices.
\end{abstract}

\section{Introduction}
\label{sec:introduction}

Magnetic confinement fusion pilot plants (FPPs) are progressing toward pilot-scale implementation, with several private efforts targeting electricity production in the coming decades. However, the expected performance of early FPP concepts still rely primarily on scaling laws and reduced modeling, which may limit optimization and reliability of extrapolation to reactor regimes. High-fidelity modeling is therefore important for advancing theoretical understanding and informing design choices, through exploratory studies of scenarios that may not yet have experimental validation.

Modeling of the boundary region of tokamaks, i.e., the region encompassing the edge and scrape-off layer (SOL),
is particularly important. This region determines the height of the H-mode pedestal that improves confinement in the core \citep{wagner1982hmode,Wagner2007}. It also affects how much confinement might be improved with low recycling liquid metal walls \citep{Krasheninnikov2008}. Furthermore, it determines whether the extreme heat fluxes to the plasma facing material can be handled \citep{Doyle2007ChapterTransport}. 
Additionally, factors such as the large amplitude of turbulent plasma fluctuations there, the complicated magnetic geometry, and the interactions of multiple physics effects, including the coexistence of microinstabilities \citep{zeiler1998}, atomic physics and plasma-wall interactions \citep{Stangeby2000}, make this region particularly challenging to model.

Among advanced modeling approaches, self-consistent turbulence evolution is a key requirement for predicting transport and can be simulated using Braginskii fluid codes such as BOUT++ \citep{dudson2009}, GBS \citep{Giacomin2022First-PrinciplesITER}, GDB \citep{zhu2018gdb,francisquez2027gdb}, GRILLIX \citep{stegmeir2018}, SOLEDGE3X \citep{bufferand2018}, and TOKAM-3D \citep{Tamain2010TOKAM-3D:Tokamaks}. These fluid frameworks can simulate edge and SOL turbulence, but their closure assumptions limit fidelity in low-collisionality regimes and in capturing kinetic microinstabilities thought to influence core and edge transport. Full-$f$ gyrokinetic (GK) codes such as COGENT \citep{dorf2020}, GENE-X \citep{ulbl2021}, \gkyl{} \citep{shi2019gkyltoklike,Hakim2022Thehttps://gkeyll.readthedocs.io}, GT5D \citep{Idomura2008GT5D}, GYSELA-X \citep{grandgirard2016}, and XGC1 \citep{chang2017}  extend fidelity by incorporating kinetic effects across a broader range of collisionalities, supporting emerging core-to-SOL studies.

The \gkyl{} computational framework includes a GK solver that implements the long-wavelength full-f GK equations in a global geometry including closed and open field lines \citep{Bernard2024PlasmaPlasmas}. A conservative discontinuous Galerkin (DG) method underpins its GK and full Vlasov solvers, conserving particles and energy independently of the resolution \citep{Juno2018DiscontinuousPlasmas}. Together with multi-GPU parallelization, this affords efficiency across a range of grid sizes. In addition to the DG GK and Vlasov solvers, \gkyl{} includes a multi-moment multi-fluid solver \citep{Juno2025pkpm}. These various solvers have been applied to problems including GK turbulence \citep{Francisquez2024ConservativeConditions,Hakim2022Thehttps://gkeyll.readthedocs.io}, magnetic reconnection, sheath physics, relativistic magneto-hydrodynamics, and astrophysical plasmas \citep{Ng2020ImprovedReconnection,Francisquez2023TowardContinuum,gorard2025hydrodynamicelectromagneticdiscrepanciesneutron,Bradshaw2025EffectsOxidation}.

Understanding and characterizing turbulence in tokamak plasmas remains a challenge with implications for future commercial fusion devices. Experiments on the Tokamak à Configuration Variable (TCV) \citep{Coda2022EnhancedTCV}, and subsequent studies on DIII-D \citep{Austin2019AchievementTokamak,Thome_2024} and ASDEX Upgrade \citep{Happel2023OverviewTokamak}, indicate that negative triangularity (NT)—a plasma shape with the triangular poloidal cross section pointing toward the tokamak's symmetry axis—can reduce turbulence levels in L-mode. Some NT discharges have reported improved H-mode-like confinement without a large pedestal \citep{Marinoni2019H-modeDIII-D,Austin2019AchievementTokamak}, potentially avoiding harmful edge localized modes \citep{Zohm1996}.

Microscale turbulence and instabilities are challenging to measure experimentally, making numerical simulations essential for understanding the interaction between triangularity and transport. Simulations with drift-Braginskii types of fluid equations successfully capture the effects of NT on turbulent transport, particularly at longer wavelengths \citep{Riva2017PlasmaTurbulence,Riva2020ShapingEffects,Lim2023EffectConfigurations,Aucone2024ExperimentsScenarios}.
\modifi{
    Beyond confinement improvements, NT configurations may also offer alternative power-exhaust mechanisms, addressing a key challenge for future fusion reactors \citep{chang2017gyrokineticdivertorheatflux,faitsch2021broadeningpowerfallofflength}.
    A strong outward shift of the NT plasma toward the low-field-side divertor can increase the heat flux deposition area through a simple geometric effect \citep{Rutherford2024MANTA}.
    However, it remains unclear whether NT is uniquely beneficial for power exhaust, as some studies report difficulties accessing detached divertor operation \citep{Tonello2024solpsNT} and a reduction of the heat flux deposition width \citep{muscente2023analysisedgetransport}.
} 
\modifi{
    Gyrokinetic modeling, which incorporates kinetic effects such as trapped electrons and finite Larmor radius (FLR) effects, provides further insights into ion gyro-radius scale physics by capturing how NT modifies the toroidal precession drift of trapped particles. This modification primarily drives the stabilization of trapped electron modes (TEM) and the reduction of electron heat transport \citep{Camenen2005ntsimTCVPlasmas,Marinoni2009ntsim}.
    More recent studies expand on these findings, demonstrating that NT influences ion temperature gradient (ITG) modes \citep{Merlo2019TurbulentTriangularity,Balestri2024PhysicalPlasmas}, and may destabilize TEM and microtearing modes (MTM) under certain conditions \citep{Hoffmann2025InvestigationSimulations,balestri2025interplaytriangularitymtm}.
    Beyond local mode analysis, global simulations reveal that NT shaping can modify the radial correlation length of turbulence and reduce the propagation of avalanche-like events \citep{DiGiannatale2024SystemSimulations}.
    However, most current GK studies focus on core-region fluctuations with prescribed boundary conditions.
    While full-$f$ simulations of the edge and SOL remain in early development, recent GENE-X results display promising outcomes in both closed and open field lines \citep{ulbl2025simulationsedgesolturbulence} but still rely on fixed profile values at the boundary, limiting predictability and requiring experimental inputs.
}

\modifi{
    In this work, we present first-of-their-kind full-$f$ flux-driven GK simulations of the edge and SOL that are free from prescribed plasma profile measurements. We validate qualitatively this framework by simulating both NT and positive triangularity (PT) discharges in the tokamak \'a configuration variable \citep{Coda2022EnhancedTCV,duval2024tcvteam}, with direct comparison to experimental measurements. 
}
We use an adaptive sourcing method to evolve edge and SOL profiles self-consistently without prescribing temperature or density boundary conditions. The total particle source rate is adjusted to maintain the desired particle inventory, modeling the experimental capability to control total particle content through gas puffing and wall conditioning.

Our simulations self-consistently evolve plasma turbulence by solving the long-wavelength full-$f$ GK equations in a global geometry that includes both closed and open field lines. This setup enables high-fidelity evolution of turbulence in the region around the last closed flux surface (LCFS). In this region, energy flows from the adaptive sources in the closed field line region to the SOL, where conducting sheath boundary conditions are applied at the limiter position. Our work leverages the implementation of twist-and-shift boundary conditions for the core region \citep{Francisquez2024ConservativeConditions} and builds upon pioneering simulations of NT discharges in DIII-D \citep{Bernard2024PlasmaPlasmas}.

With the recent implementation of an adaptive sourcing scheme, \gkyl{} can in these cases be operated using three inputs: magnetic equilibrium, heating power, and total plasma mass. Under these assumptions it determines quasi-stationary turbulent profiles without prescribing boundary conditions from experimental measurements. The intent behind this capability, while still subject to model assumptions (e.g. simplified recycling treatment, electrostatic limit), is twofold: first, to enable validation against experimental measurements with a minimum of experimentally inferred inputs; second, to support studies of scenarios where experimental profile data are unavailable, e.g. for future fusion power plant design optimization.
Furthermore, present efficiency and scalability allow simulations of milliseconds of medium-sized tokamak turbulence with kinetic electrons and ions in a few hundred GPU hours, which may be sufficient for exploratory parameter studies.

The remainder of this paper is organized as follows. Section~\ref{sec:model} summarizes the GK equations and numerical model employed in these \gkyl{} simulations. Section~\ref{sec:adaptive_sourcing} details the adaptive sourcing technique. Section~\ref{sec:simulation_setup} presents the simulation setup for PT and NT TCV discharges. Section~\ref{sec:results} presents validation results for a PT simulation against experimental measurements and compares PT and NT simulations to investigate triangularity effects on turbulent transport. Finally, Section~\ref{sec:conclusions} summarizes findings and outlines potential extensions.

\section{Gyrokinetic model}
\label{sec:model}

In this work, we consider a magnetized plasma consisting of two species: deuterium ions and electrons. The plasma is described by the full-$f$ GK distribution function $f_s(\mathbf{R}, v_{\parallel}, \mu, t)$ for each species $s$, where $\mathbf{R}$ is the guiding-center position, $v_{\parallel}$ is the velocity parallel to the magnetic field $\bm B$, $\mu = m_s v_{\perp}^2/(2B)$ is the magnetic moment (an adiabatic invariant) with $v_{\perp}$ the perpendicular velocity, $m_s$ the mass of species $s$, $B$ the magnetic field amplitude, and $t$ is time.
The \gkyl{} code evolves these distribution functions in the electrostatic, long-wavelength limit, $k_{\perp}\rho_i \ll 1$, where $k_{\perp}$ is the perpendicular wave number and $\rho_i$ is the ion Larmor radius. This approximation neglects finite Larmor radius (FLR) effects in the gyrokinetic (GK) equation itself, while retaining lowest-order FLR contributions in the quasineutrality condition through the ion polarization term.

\subsection{Gyrokinetic Equation}

The full-f long wavelength GK equation solved in \gkyl{} is \citep{Francisquez2025velocitymappings},

\begin{equation}
    \label{eq:gk}
    \frac{\partial B_\parallel^* f_s}{\partial t} + \nabla \cdot (B_\parallel^* \{\mathbf{R}, H\} f_s) + \frac{\partial}{\partial v_{\parallel}}(B_\parallel^* \{v_{\parallel}, H\} f_s) = B_\parallel^* C[f_s] + B_\parallel^* S_s,
\end{equation}
where $H = \frac{1}{2}m_s v_{\parallel}^2 + \mu B + q_s \phi$ is the Hamiltonian, with $\phi$ the electrostatic potential and $q_s$ the charge of species $s$,
\begin{equation}
    \{F, G\} = \frac{\bm B^*}{m_s B^*_{\parallel}} \cdot \left(\nabla F \frac{\partial G}{\partial v_{\parallel}} - \nabla G \frac{\partial F}{\partial v_{\parallel}}\right) - \frac{1}{q_s B^*_{\parallel}} \mathbf{b} \cdot (\nabla F \times \nabla G),
    \label{eq:poisson_bracket}
\end{equation}
is the Poisson bracket, $C[f_s]$ is the collision operator, and $S_s$ is a source term.
In Eqs.~\eqref{eq:gk} and \eqref{eq:poisson_bracket}, we introduce the Jacobian of the guiding center coordinate, $B^*_{\parallel} = \mathbf{b} \cdot \mathbf{B}^*$, which is also the parallel component of the effective magnetic field, $\mathbf{B}^* = \mathbf{B} + (m_s v_{\parallel}/q_s) \nabla \times \mathbf{b}$, with $\mathbf{b} = \mathbf{B}/B$ the unit vector along the magnetic field.

\subsection{Quasineutrality Equation}
The system is closed with the long wavelength GK quasineutrality equation,
\begin{equation}
    -\sum_s n_{s0}\frac{q_s^2\rho^2_{s0}}{T_{s0}}\nabla_{\perp}^2\phi = \sum_s q_s n^g_s(\mathbf{R}, t),
    \label{eq:quasineutrality}
\end{equation}
where $n_{s0}$ is the reference density of species $s$, $\rho_{s0}=\sqrt{m_s T_{s0}}/(e B_0)$ its reference Larmor radius, $T_{s0}$ its reference temperature, $B_0$ the reference magnetic field, $e$ the elementary charge, and $n^g_s$ the guiding-center density of species $s$,
\begin{equation}
    n^g_s(\mathbf{R}, t) = \int d v_{\parallel} d\mu \, 2\pi B_\parallel^* f_s(\mathbf{R}, v_{\parallel}, \mu, t).
\end{equation}
Equation \eqref{eq:quasineutrality} retains lowest-order FLR effects present in the polarization term on the left-hand side, assuming Boussinesq approximation.

\subsection{Collision Operator}

Collisions are modeled using a multiple species version of the long wavelength GK Lenard-Bernstein-Dougherty (LBD) collision operator \citep{Lenard1958,Dougherty1964,Francisquez2022improvedDougherty},
\begin{equation}
    J_s C[f_s] = \sum_r \nu_{sr} \nabla_v \cdot \left[ \left( \bm v - \bm u_{rs} \right) J_s f_s + v_{t,sr}^2 \nabla_v J_s f_s \right],
\end{equation}
where $\nu_{sr}$ is the collision frequency of species $s$ with species $r$, $\bm u_{rs}$ and $v_{t,rs}$ are the cross-species primitive moments designed to enforce exact momentum and energy conservation \citep{Francisquez2022improvedDougherty}.
In this work, the collision frequencies are time and space dependent, $\nu_{sr}=\nu_{sr}(\bm R,t)$, and are computed as
\begin{equation}
    \nu_{sr}(\bm R,t) = \nu_{sr0} \frac{n_r(\bm R,t)}{n_{r0}} \left[\frac{v_{ts0}^2 + v_{tr0}^2}{v_{ts}^2(\bm R,t) + v_{tr}^2(\bm R,t)}\right]^{3/2},
\end{equation}
where
\begin{equation}
    \nu_{sr0} = \frac{1}{m_s}\left(\frac{1}{m_s}+\frac{1}{m_r}\right)\frac{q_s^2q_r^2\log\Lambda_{sr}}{3(2\pi)^{3/2}\epsilon_0^2}\frac{n_{r0}}{\left(v_{ts0}^2+v_{tr0}^2\right)^{3/2}},
\end{equation}
with $\log\Lambda_{sr}\sim 14$ the Coulomb logarithm.

\subsection{Geometry and Coordinate System}
Our \gkyl{} simulations use a field-line-following coordinate system defined by
\begin{align}
    x &= r - a, \label{eq:mapc2p_x}\\
    y &= -\frac{r_0}{q_0}[\varphi + \alpha(r, \theta, \varphi = 0)], \label{eq:mapc2p_y} \\
    z &= \theta \label{eq:mapc2p_z},
\end{align}
where $x$ is the radial coordinate, $y$ is the bi-normal coordinate, and $z$ is the parallel coordinate.
In Eqs. \ref{eq:mapc2p_x}--\ref{eq:mapc2p_z},  $r$ is the minor radius coordinate, $\theta$ the poloidal angle, $\varphi$ the toroidal angle, $r_0$ the reference radius (usually the radial center of the domain), $q_0=q(r_0)$ the reference safety factor, $a$ the LCFS minor radius, and $\alpha(r, \theta, \varphi)$ the field-line label function,
\begin{equation}
    \alpha(r, \theta, \varphi) = \varphi - 2\pi q(r) \frac{\int_0^\theta J_r(r, \theta')/R(r, \theta')^2 d\theta'}{\int_0^{2\pi} J_r(r, \theta')/R(r, \theta')^2 d\theta'},
\end{equation}
where $J_r$ is the configuration space coordinate Jacobian.
The non-orthogonal field-line-following coordinates $(x, y, z)$ are defined such that
\begin{equation}
    \mathbf{B} = \frac{J_r B}{\sqrt{g_{zz}}}(\nabla x \times \nabla y),
\end{equation}
where $g_{zz}$ is the covariant metric tensor component. The effect of the other metric tensor components, $g_{xx}$, $g_{yy}$, and $g_{xy}$, are also included indirectly through the curvilinear coordinate system used to represent our computational domain \citep{Francisquez2025velocitymappings}.

The equilibrium magnetic flux surfaces are described by a version of the Miller parameterization \citep{Miller1998NoncircularModel} where the Shafranov shift is assumed to vary quadratically with minor radius,
\begin{align}
    R(r, \theta) &= R_\text{axis} - \alpha_s \frac{r^2}{2 R_\text{axis}} + r \cos(\theta + \sin^{-1}\delta \sin\theta), \\
    Z(r, \theta) &= Z_\text{axis} + r \kappa \sin\theta,
\end{align}
where $R_\text{axis}$ is the major radius of the magnetic axis, $Z_\text{axis}$ its vertical position, $\delta$ the triangularity, $\kappa$ the elongation, and $\alpha_s$ the Shafranov shift parameter.
The LCFS minor radius, $a$, is obtained self-consistently by solving the second order equation
$R(a, 0) = R_\text{LCFS}$, where $R_\text{LCFS}$ is the LCFS major radius at the outboard midplane (OMP), which yields,
\begin{equation}
    a = \frac{R_\text{axis}}{\alpha_s}
    \left(
        1 - \sqrt{1 - 2\alpha_s 
        \epsilon}
    \right),
\end{equation}
where $\epsilon =(R_\text{axis}-R_\text{LCFS})/R_\text{axis}$ is the inverse aspect ratio.
This Miller geometry model, valid for small inverse aspect ratio, does not retain squareness and higher-order shape radial derivatives \citep{Turnbull1999}; these effects are neglected here.

\subsection{Boundary Conditions}

The simulation domain includes both closed and open field line regions, requiring different boundary condition treatments in each region.
In the radial direction, we assume a zero distribution function outside the simulation domain.
Consequently, particles that are drifting radially toward both the core and the wall are absorbed in the inner and outer radial boundaries, respectively, and no particles can enter the simulation domain from the radial boundaries.
For the electrostatic potential, Dirichlet boundary conditions ($\phi = 0$ V) are applied, which eliminates the $E\times B$ drift and implies that the loss of particles through the radial boundaries is solely due to the $\grad B$ drift.

In the bi-normal direction, periodic boundary conditions are assumed for both the distribution functions and the potential, following a standard flux-tube approach. This approximation is expected to hold so long as the bi-normal extent exceeds the turbulence correlation length.

In the parallel direction, two types of boundary conditions are applied depending on the closed or open field line region.
For closed flux surfaces, twist-and-shift boundary conditions connect the twisted ends of the flux tube domain \citep{Beer1995,Ball2020,Francisquez2024ConservativeConditions},
\begin{equation}
    F(x, y(z + L_z), z + L_z) = F(x, y(z), z),
\end{equation}
where $F$ represents any field quantity and $L_z$ is the parallel domain length.
From the field-line label function $\alpha(r, \theta, \varphi = 0)$, we can derive,
\begin{equation}
    \alpha(r,z,\varphi = 0) = -2\pi q C \int_{0}^{z} \frac{J}{R^2} dz' \quad \text{with} \quad C^{-1} = \int_{0}^{2\pi} \frac{J}{R^2} dz',
\end{equation}
which leads to the $y$-coordinate transformation,
\begin{equation}
    y(z+\Delta z) = y(z) + \frac{r_0}{q_0} 2\pi q C \int_{z}^{z+\Delta z} \frac{J}{R^2} dz'.
\end{equation}
In our simulations, the twist-shift boundary conditions are applied at $z = \pm\pi$ rather than at the full $2\pi$ domain. For a Miller geometry with up-down symmetry, the matching condition becomes,
\begin{equation}
    F(x, y_0 + S_y(r,\pi) , \pi) = F(x, y_0 + S_y(r,-\pi), -\pi),
\end{equation}
with the shift function
\begin{align}
    S_y(r,z) &= \frac{r_0}{q_0} \alpha(r, z, \varphi=0),
\end{align}
In the up-down symmetric case, the shift functions simplifies to $S_y(r,\pi) = +\pi r_0 q(r)/q_0$ and $S_y(r,-\pi) = -\pi r_0 q(r)/q_0$.

In the open field line region, conducting sheath boundary conditions are applied to the electrostatic potential, which self-consistently builds up a sheath potential.
Combined with a zero distribution function inside the limiter, the sheath boundary condition reflects low-energy electrons while allowing ions and high-energy electrons to flow freely out of the domain \citep{Shi_Hammett_Stoltzfus-Dueck_Hakim_2017}.

To conserve energy with our DG scheme, the potential must be continuous \modifi{\citep{Francisquez2024ConservativeConditions}}. For this reason we apply the twist-and-shift boundary condition on $\phi(z=-\pi)$ in the closed field line region, so that it is twist-shift periodic relative to the upper end of the domain (at $z=\pi$).
Additionally, continuity must be ensured at the limiter edge, i.e. $x=x_{LCFS}$ and $z=\pm \pi$, where twist-and-shift and sheath boundary conditions are touching.
This is achieved by applying a Dirichlet boundary condition at the limiter edge, i.e. $\phi(x_{LCFS}, y, z=\pm \pi) = 0$ V for all $y$. 
While the limiter corner could be treated more rigorously, this simplified approach appears sufficient for the present work. We report that applying a different bias to the limiter corner region does not strongly affect the bulk turbulence dynamics in the simulated SOL.

\subsection{Adaptive Sourcing}
\label{sec:adaptive_sourcing}

\modifi{A key new feature of this work is the implementation of an adaptive sourcing scheme in \gkyl{}, which dynamically adjusts the particle and energy fluxes of the source in response to boundary losses.}
This approach relies on the ability of \gkyl{} to measure particle and energy fluxes through the domain boundaries at each Runge-Kutta sub-step. This is natural in a discontinuous Galerkin framework, as the fluxes must be computed as part of the numerical scheme.\\
We express the total volumetric source term in the gyrokinetic equation (Eq.~\ref{eq:gk}) for species $s$ as a sum of $N_S$ sub-sources,
\begin{equation}
    S_s = \sum_{k=1}^{N_S} S^k_s.
\end{equation}
Each sub-source $S^k_s$ is a non-drifting Maxwellian in velocity space, and Gaussian in configuration space, i.e.,
\begin{equation}
    S^k_s\left[\mathbf{R}, v_{\parallel}, \mu; \Gamma^k_s, T^k_s\right] = G^k_s(\bm R; \Gamma^k_s)\, M^k_s\left[v_{\parallel}, \mu; T^k_s\right],
\end{equation}
where we separate by a semicolon the coordinate dependence and the parameters of the source, i.e. the particle injection rate $\Gamma^k_s$ and temperature $T^k_s$.
The Gaussian configuration space profile is defined as,
\begin{equation}
    G^k_s(\bm R; \Gamma^k_s) = \Gamma^k_s \eta^k_s(\bm R),
\end{equation}
with a normalized spatial profile,
\begin{equation}
    \eta^k_s(\mathbf{R}) = C^k_s \prod_{i=x,z} \exp\left[-\frac{(x_i - \xi_{i,s,k})^2}{2\sigma_{i,s,k}^2}\right],
\end{equation}
where $C^k_s$ ensures $\int \eta^k_s d\mathbf{R}= 1$.
The Maxwellian velocity space profile is defined as,
\begin{equation}
    M^k_s\left[v_{\parallel}, \mu; T^k_s\right] = \left(\frac{m_s}{2\pi T^k_s}\right)^{3/2} \exp\left[-\frac{m_s v_{\parallel}^2}{2T_s^k}-\frac{\mu B}{T^k_s}\right],
    \label{eq:maxwellian_shape}
\end{equation}
where the flow velocity is neglected and can be included in future work.

Given a target input particle rate and power, $\Gamma_{k,s}^\text{in}$ and $P_{k,s}^\text{in}$, respectively, the adaptive sub-source parameters also compensate for instantaneous particle and power losses of a species $r$, through a given set of boundaries, $\dot N^k_r$ and $\dot E^k_r$, respectively.
The update of the sub-source parameters is done at each time step, $t_n$, as follows,
\begin{align}
    \Gamma^k_s(t_{n+1}) &= \Gamma_{k,s}^\text{in} + \dot N^k_r(t_n), \\
    T^k_s(t_{n+1}) &= \frac{2}{3} \frac{P_{k,s}^\text{in} + \dot E^k_r(t_n)}{\Gamma^k_s(t_{n+1})}.
\end{align}
We note that the source for a species $s$ can compensate for losses of a different species $r$, e.g. an ion source compensating for electron losses, to model ambipolar processes such as recycling.
This sub-source adaptation allows construction of sources intended to emulate selected processes in the plasma.
In the next section, we present a source setup composed of two adaptive sub-sources per species representing core-to-edge heat transport and plasma–wall recycling in a simplified manner.

\section{Simulation Setup}
\label{sec:simulation_setup}
In this section, we describe the TCV plasma parameters and the numerical setup.
We consider two TCV L-mode deuterium plasma discharges, \#65125 (PT) and \#65130 (NT), with times of interest 0.5~s and 1.3~s, respectively. 
Both use Ohmic power as the main heating source, and the SOL power $P_\text{SOL}$---i.e., the power measured neglecting radiative losses and stored power---is approximately $0.38$~MW for PT and $0.35$~MW for NT.

\subsection{Geometry and Plasma Parameters}

For plasma parameters, we consider deuterium ions with mass $m_i = 2.01\,m_p$ (where $m_p$ is the proton mass) and kinetic electrons with realistic ion-electron mass ratio, $m_e/m_i=2.7\times 10^{-4}$. 
The reference temperature for both species is set to $T_{e0} = T_{i0} = 100$ eV, with a reference density of $n_0 = 2\times10^{19}$ m$^{-3}$.
The reference magnetic field amplitude is set to the value at the center of the OMP, i.e. $B_{0} = 1.13$~T.
Consequently, the reference ion sound speed is $c_{s0} = \sqrt{T_{e0}/m_i} \simeq 6.9\times 10^4$~m/s, the reference sound Larmor radius is $\rho_{s0} \simeq 1.3\times 10^{-3}$~m, the sound wave propagation time is $R_\text{axis}/c_{s0} \simeq 1.0\times 10^{-5} $~s.
We report reference collision times and frequencies in Tab.~\ref{tab:collision_freq}.

\begin{table}
\centering
\begin{tabular}{lcccc}
    Species $s$-$r$ & $\nu_{sr}$ [s$^{-1}$] & $\tau_{sr}$ [s] & $\tau_{sr} c_{s0}/R$ & $\nu_{sr}^*$ \\
    elc-ion & $1.80 \times 10^{6}$ & $5.55 \times 10^{-7}$ & $4.31 \times 10^{-2}$ & $1.59 \times 10^{1}$ \\
    elc-elc & $6.23 \times 10^{5}$ & $1.60 \times 10^{-6}$ & $1.25 \times 10^{-1}$ & $5.49 \times 10^{0}$ \\
    ion-ion & $1.04 \times 10^{4}$ & $9.61 \times 10^{-5}$ & $7.47 \times 10^{0}$ & $5.58 \times 10^{0}$ \\
    ion-elc & $4.63 \times 10^{2}$ & $2.16 \times 10^{-3}$ & $1.68 \times 10^{2}$ & $2.48 \times 10^{-1}$ \\
\end{tabular}
\caption{Reference collision frequencies, times, and normalized collision parameters for the PT simulation. The normalized collision parameter $\nu^*$ is defined as $\nu^* = \nu_{sr}q_0 R_0/\epsilon^{3/2} v_{ts}$.}
\label{tab:collision_freq}
\end{table}

%
\begin{figure}
    \centering
    \includegraphics[width=0.49\linewidth]{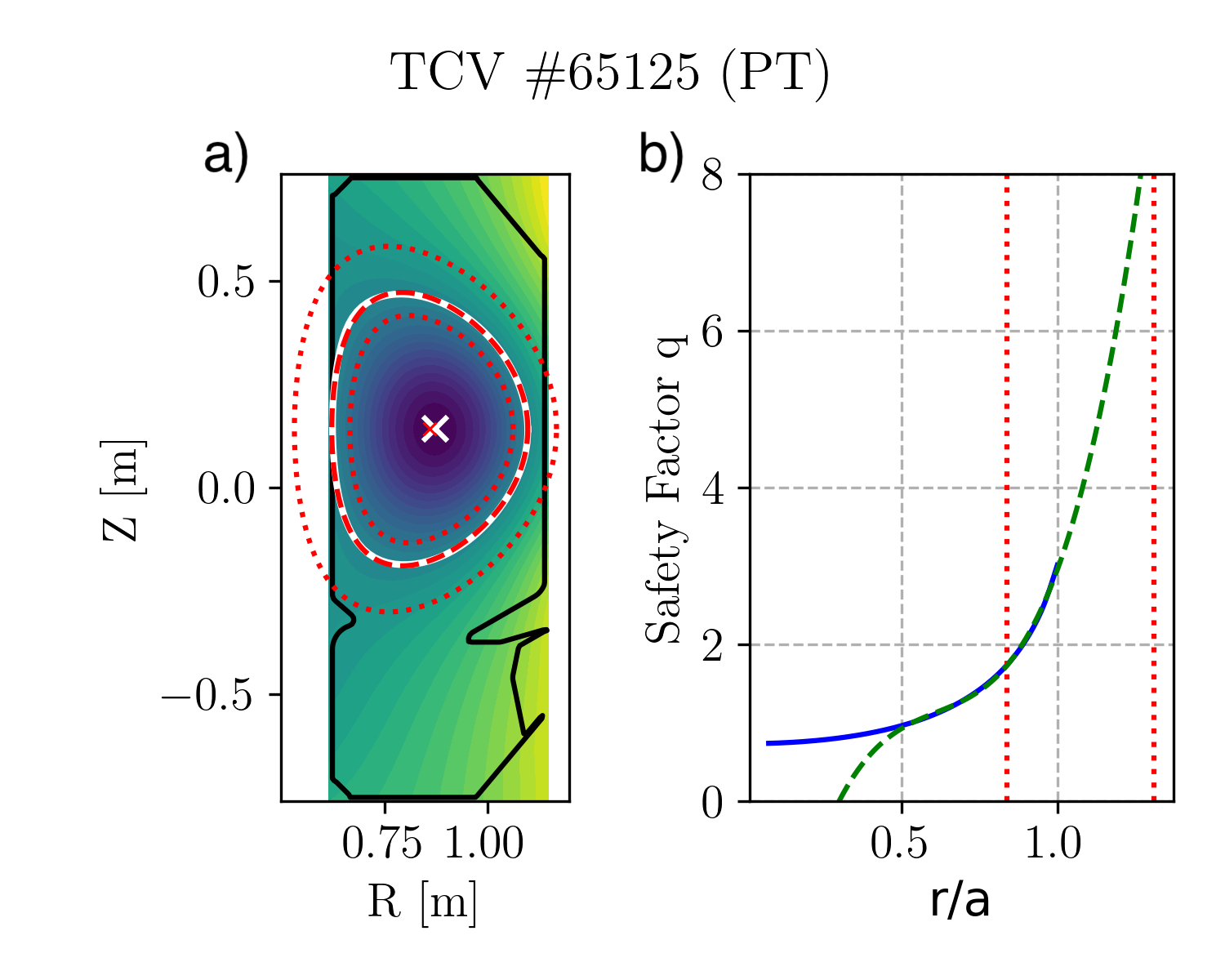} 
    \includegraphics[width=0.49\linewidth]{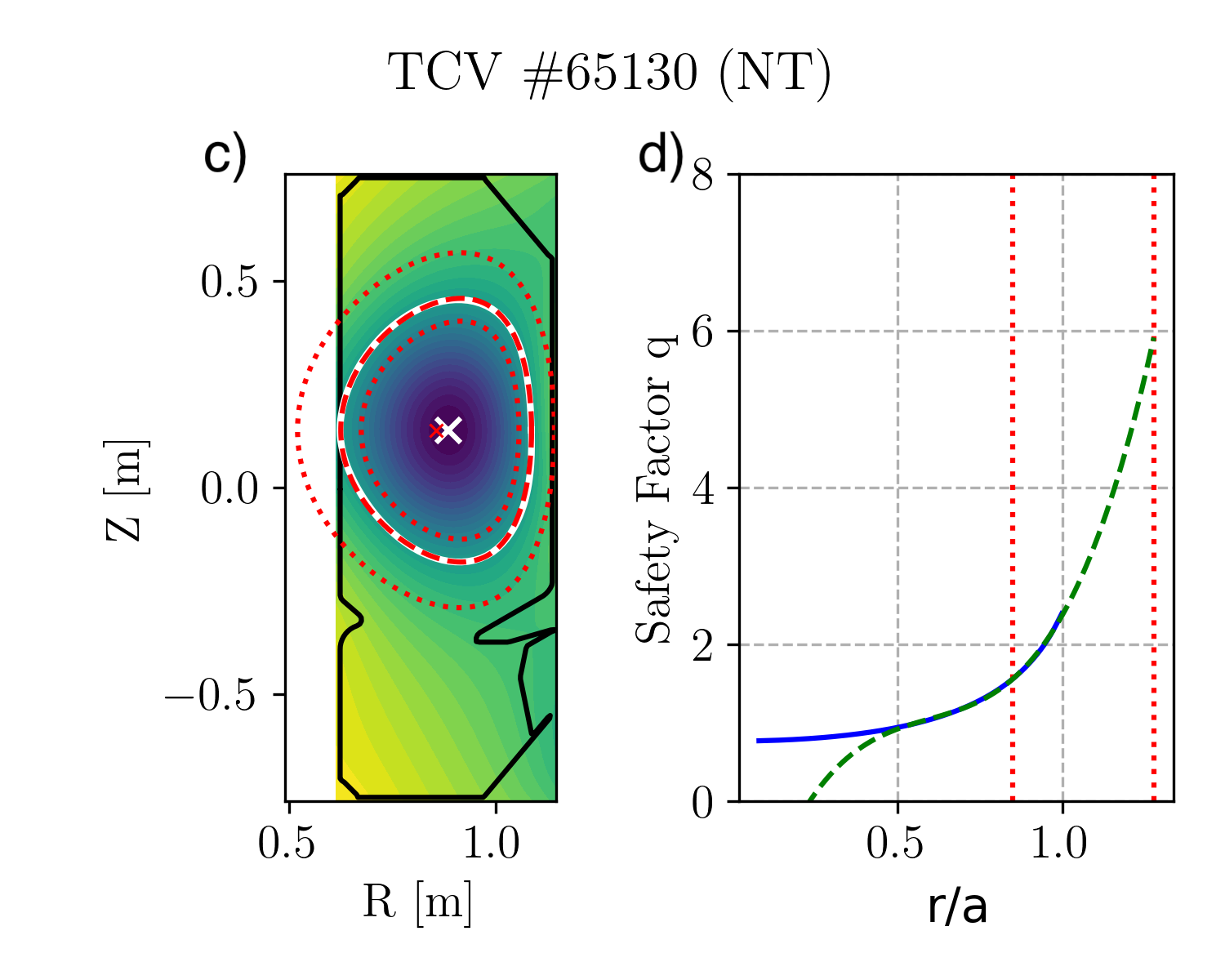} 
    \caption{TCV magnetic equilibrium for PT (a and b) and NT (c and d) discharges. Panels (a) and (c) show poloidal cross-sections of the TCV vessel with poloidal flux contours (color scale), vessel wall (solid black line), LCFS (solid white line), and magnetic axis (white "x"). Miller equilibrium parameters used in \gkyl{} are overlaid in red: LCFS (dashed line), maximal and minimal radii (dotted lines), and shifted axis ("x"). Panels (b) and (d) show the reconstructed safety factor profile (solid blue line) and the cubic fit used in simulations (dashed green line). We also report the radial limits of the simulation domain (dotted red lines).}
    \label{fig:tcv_miller_equilibrium}
\end{figure}


For each discharge, we set the flux tube geometry by fitting the Miller equilibrium model parameters (such as elongation, triangularity, and Shafranov shift) to the magnetic equilibrium reconstruction data at the LCFS, obtained from EFIT, using the least squares method.
The safety factor profile $q(R)$ is obtained by applying a cubic fit to the magnetic equilibrium reconstruction data for \modifi{$r/a \geq 0.6$}, where the cutoff is chosen to focus on the edge region and avoid core-specific features that are not relevant for the present work.
The key geometry parameters, such as elongation, triangularity, and Shafranov shift, are summarized in Tab.~\ref{tab:tcv_geom_param} for both PT and NT discharges.
Aside from the triangularity---which is positive for PT and negative for NT---most of the shaping parameters are kept constant between the two configurations.
We note however a larger Shafranov shift in the NT discharge, and a larger safety factor in the PT discharge. 
The resulting magnetic equilibria are illustrated in Fig.~\ref{fig:tcv_miller_equilibrium}.
\modifi{For the rest of this manuscript, we use the normalized minor radius, $r/a$, to refer to radial locations at the OMP, where $r/a=1$ corresponds to the LCFS/separatrix location.}

\begin{table}
    \centering
    \begin{tabular}{lcc}
         & PT & NT \\
        Magnetic axis, $R_{\text{axis}}$ [m] & 0.87 & 0.89 \\
        Magnetic axis, $Z_{\text{axis}}$ [m] & 0.14 & 0.16 \\
        Axis field, $B_{\text{axis}}$ [T] &  1.40 & 1.40 \\
        LCFS (OMP), $R_{\text{LCFS}}$ [m] & 1.10 & 1.09 \\
        Shafranov shift, $\alpha_s$ & 0.25 & 0.5 \\
        Elongation, $\kappa$ & 1.45 & 1.40 \\
        Triangularity, $\delta$ & 0.35 & $-0.38$ \\
        Safety factor fit coeff. & $[497.34, -1408.74, $ & $[484.06, -1378.26,]$ \\
        $q(R)=\sum_{n=0}^3 a_{3-n} R^n$ & $1331.41, -419.01]$ & $1309.31, -414.13]$ \\
    \end{tabular}
    \caption{Magnetic geometry and safety factor parameters for TCV Miller equilibria. PT refers to the discharge \#65125, and NT refers to the discharge \#65130.}
    \label{tab:tcv_geom_param}
\end{table}

The simulation domain used in \gkyl{} covers both closed and open field line regions, enabling study of edge and scrape-off layer (SOL) dynamics with turbulence seeded by first-principles edge GK microinstability.
The flux tube domain extends radially 4 cm inside the LCFS and 8 cm into the SOL, yielding $L_x\simeq 100\rho_{s0}$. 
The bi-normal length is set to $L_{y}=150\rho_{s0}$, yielding $L_y \simeq 14$ cm.
In the parallel direction, the flux tube wraps toroidally to cover one poloidal turn, which is sufficient at high magnetic shear, where parallel correlation length is small.

\subsection{Source Setup}
\label{sec:source_setup}
The source setup consists of two main sub-sources: the \textit{core source}, $S^\text{core}_s$, and the \textit{recycling source}, $S^\text{recy}_s$, which are described in detail below.
The core source is designed to inject heating power transported from the tokamak core— not simulated here—without net particle addition, consistent with negligible neutral beam injection (NBI) heating for these discharges.
While pure power injection can be achieved via an antenna and electromagnetic fluctuations \citep{Ohana_2018}, we leverage here the particle losses induced by the magnetic gradient-B drifts at the inner radial boundary.
We use the reinjection of lost particles at the $x=0$ boundary to add a net power $P_{\text{inj}}=v_f \times P_{\text{exp}}$, that is the experimental heating power scaled by the volume fraction of the flux tube, $v_f = q_{0}L_y/2\pi r_0 \simeq 0.4$, and split evenly between the two species.
In the considered TCV magnetic equilibrium, the ion gradient-B drift is directed downwards (negative $Z$ direction), while the electron gradient-B drift is directed upwards.
Hence, we center the ion core source at $z=-\pi/2$ (i.e., $\xi_{z,i,c}=-\pi/2$) and the electron core source at $z=\pi/2$, to mimic the spatial separation of particle sources due to magnetic gradient-B drift directions.

\begin{figure}
    \centering
    \includegraphics[scale=0.16]{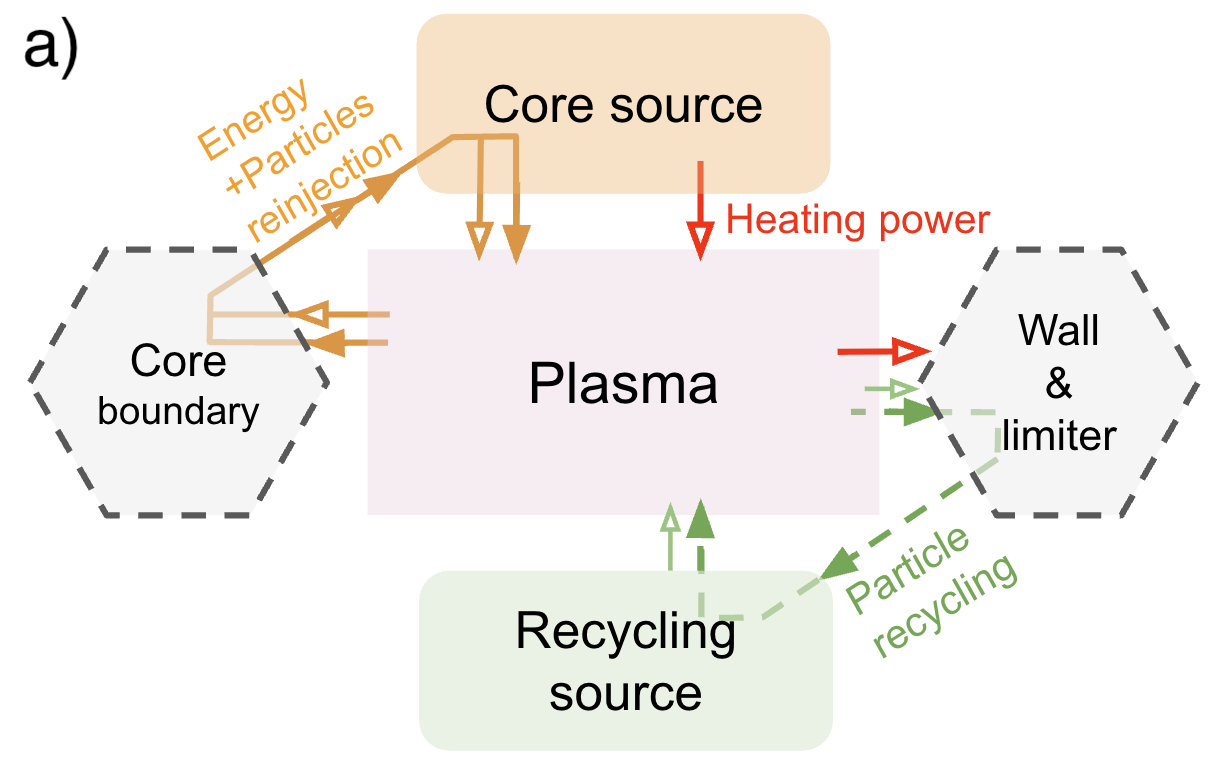} 
    \includegraphics[scale=0.16]{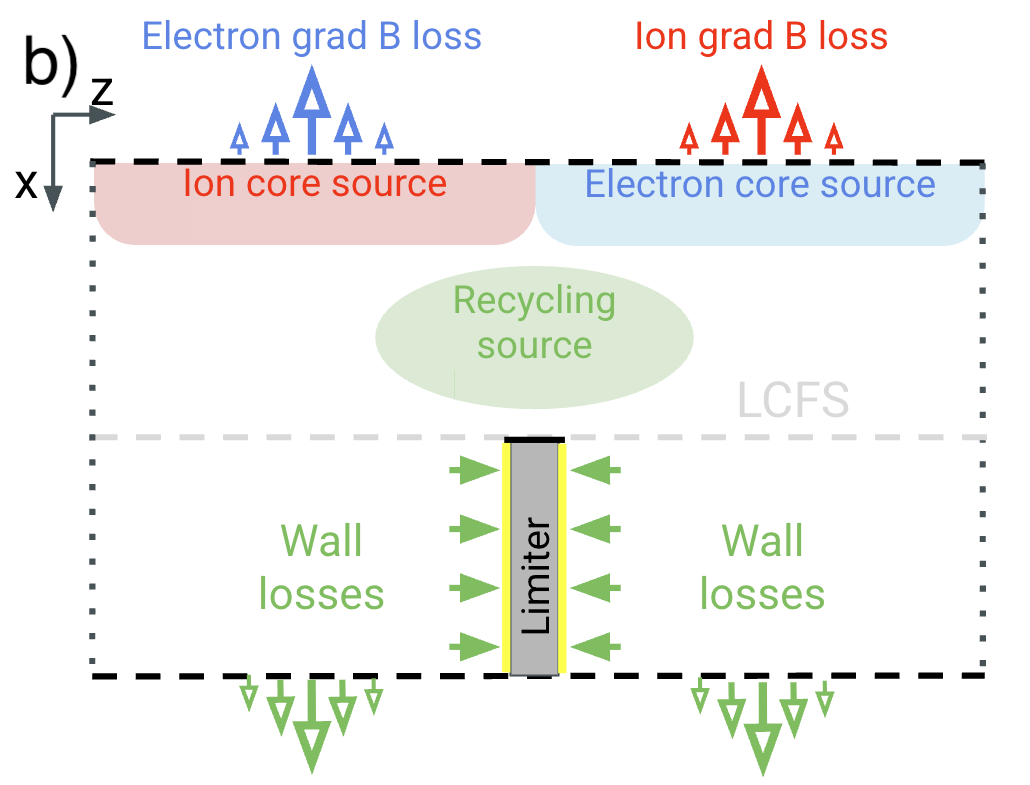} 
    \caption{
    Setup of the adaptive sources. 
    The left panel illustrates the particle fluxes (solid arrows), energy fluxes (hollow arrows) and feedback loops for adjusting source parameters based on measured losses.
    The right panel illustrates the positions of the core and recycling sources, with the core source centered at the inner radial boundary and the recycling source near the limiter.
     }
    \label{fig:source_setup}
\end{figure}

The recycling source emulates ionization of neutrals generated by ions lost to the limiter surface; these ions subsequently return as neutrals and are re-ionized in the plasma. 
The particle rate of this source is dynamically adjusted to compensate for ion losses through the outer radial boundary and limiter, injecting equal numbers of ions and electrons with a temperature of 10~eV.
This source contributes marginal power injection into our system, representing less than 1\% of the total power input.
The shape and position of the recycling source is inspired by \cite{Coroado2022ABoundary} where a similar discharge is studied with the GBS code using a kinetic model for neutrals.
While the recycling source reproduces the gross density injection on the high-field side, it does not capture localized cooling associated with ionization nearer the limiter and does not model neutral transport explicitly. This approximation avoids introducing prescribed neutral density profiles. For more comprehensive predictive capability, the recycling source would need replacement by a coupled neutral model, which is beyond the present scope and would increase computational cost significantly.
The interactions between the sources and the losses are illustrated in Fig.~\ref{fig:source_setup}. Table~\ref{tab:source-param} summarizes the parameters for both the core and recycling sources used in this study.

\begin{table}
    \centering
    \begin{tabular}{lcc}
        & Core Source & Recycling Source \\
        $\dot E^k_r$ & $r=s$ at $x=0$ & none   \\
         $\dot N^k_r$ & $r=s$ at $x=0$  & $r=i$ at $x=L_x$ \& $z=\pm \pi$   \\
        $P^{in}_{k,s}$ & $v_f P_{exp}/2$ & $0$ \\
        $\Gamma^{in}_{k,s}$ & $0$  & $0$  \\
        $(\xi_{x,i,k},\xi_{z,i,k})$ & $(x_\text{min},\ -L_z/4)$ & $(3x_\text{LCFS}/4,\ \pi)$ \\
        $(\xi_{x,e,k},\xi_{z,e,k})$ & $(x_\text{min},\ L_z/4)$ & $(3x_\text{LCFS}/4,\ \pi)$ \\
        $(\sigma_{x,s,k},\sigma_{z,s,k})$ & $(3L_x/100,\ L_z/6)$ & $(\,x_\text{LCFS}/4,\ L_z/20)$ \\
    \end{tabular}
    \caption{Parameters for the core and recycling sources. Note that the recycling source compensate for ion losses for both species, to ensure ambipolarity.}
    \label{tab:source-param}
\end{table}

\subsection{Initial Conditions}
The simulations are initialized with radially varying profiles of density to approximately match the experimental TCV profiles in order to obtain a comparable particle number,
\begin{equation}
    n(x) = n_0 \left(1.0 + \tanh\left[-80\left(x - x_0\right) \right] \right) + 0.001 n_0
\end{equation}
where $n_0 = 2\times10^{19}$ m$^{-3}$ is the reference density, and the transition region is centered at $x_0 = -0.03$ m, i.e. near the LCFS position. 
This initial condition is used for both PT and NT simulations, as the variation in the geometry does not significantly affect the total volume of the flux tube -- the total number of particle is $N \simeq 1.5\times 10^{19}$ for PT and $N \simeq 1.4\times 10^{19}$ for NT.

The initial electron and ion temperatures profile are set to
\begin{equation}
    T_e(x) = T_{e0} \left(0.8 + 0.7\tanh[-80(x - x_0)]\right),
\end{equation}
and
\begin{equation}
    T_i(x) = T_{i0} \left(0.7 + 0.5\tanh[-30(x - x_0)]\right),
\end{equation}
respectively. 
Aside from the total particle inventory, the steady-state profiles appear insensitive to the chosen initial shapes. The initial conditions presented here are selected to reduce the transient duration.

\subsection{Numerical Resolution}

The \gkyl{} simulations presented in this work use a discontinuous Galerkin (DG) method with polynomial order $p=1$ in configuration space and a hybrid linear-quadratic basis in phase space, yielding 48 degrees of freedom per cell.
The configuration space is discretized using a uniform grid in the radial, $x$, bi-normal, $y$, and parallel, $z$, directions with $N_x$, $N_y$, and $N_z$ cells, respectively.
We consider different numerical resolutions in configuration space. 
The coarse resolution uses $(N_x, N_y, N_z) = (24, 16, 12)$, the baseline resolution uses $(N_x, N_y, N_z) = (48, 32, 16)$, and the fine resolution uses $(N_x, N_y, N_z) = (96, 64, 16)$ cells. 
In velocity space, non-uniform grids allow efficient resolution of the thermal population while still capturing suprathermal particles \citep{Francisquez2025velocitymappings}.
The parallel velocity domain spans $-6 \leq v_\parallel/v_{ts} = 6$ with $v_{ts}=\sqrt{T_{s0}/m_s}$ the thermal speed, and is discretized using a linear mapping up to $v_\parallel=2v_{ts}$, for $s=i,e$, using $N_{v\parallel}=12$ cells. 
Magnetic moments are defined for $\mu \leq 1.5m_e(4v_{ts})^2/(2B_0)$, for $s=i,e$, with a quadratic mapping and $N_\mu=8$ cells.

\modifi{
    While the spatial resolutions used in these simulations may appear lower than those typically employed in local flux-tube gyrokinetic simulations, several factors must be considered. 
    First, our DG scheme encodes two basis elements per dimension (three in $\vpar$), resulting in 48 degrees of freedom per cell in phase space \citep{Shi_Hammett_Stoltzfus-Dueck_Hakim_2017}.
    The equivalent finite difference resolution of our fine resolution case therefore corresponds to $(N_x, N_y, N_z) = (192, 128, 64)$ and $(N_{v\parallel},N_\mu) = (36,16)$, which is comparable to typical flux-tube GK simulations.
    Second, the radial extent considered here is significantly larger than typical flux-tube simulations, resulting in a coarser effective resolution per unit length. However, this is well-suited for edge and SOL turbulence, which is characterized by longer wavelength structures.
    Finally, our simulations naturally limit small-scale turbulence excitation through profile relaxation, high collisionality, and the generation of large-scale $\ExB$ shear flow, which reduce the microinstability drive and suppress small-scale turbulence \citep{Hammett2006ImplementationCode,Hoffmann2023GyrokineticOperators}, further reducing the resolution requirements.
}

\section{Results}
\label{sec:results}
In this section, we present results from flux-driven gyrokinetic simulations of TCV discharges. 
The simulations evolve plasma turbulence for 2~ms, approaching a quasi-steady state and resolving tens of ion–ion collision times and hundreds of sound-wave propagation times (Tab.~\ref{tab:collision_freq}).

In terms of computational cost, the coarse resolution simulation reaches $1$~ms in roughly 40 wall-clock hours using one Perlmutter node (4 NVIDIA A100-PCIE-40GB GPUs), i.e. about 160 GPUh/ms. For the baseline resolution, the rate is about 750 GPUh/ms on 8 GPUs, and the high resolution case reaches 1 ms in approximately 2000 GPUh using 16 GPUs. The simulations use an average time step of $\sim 1$~ns, limited here by the collision frequency; without collisions, the explicit time step would increase to $\sim 6$~ns. Ongoing work on implicit treatment of collisions (e.g. with a Krook operator \citep{Liu2025collision}) may alleviate this constraint.
The input parameters for the simulations and details about the version of \gkyl{} used are provided at \url{https://github.com/ammarhakim/gkyl-paper-inp/tree/master/2025_NF_tcvadaptsrc}.

\subsection{Plasma simulation of TCV \#65125 discharge}
\label{sec:results_pt}
In this subsection we examine in detail the simulation of TCV \#65125 discharge (PT).
First, we study convergence of the quasi-steady state kinetic profiles and compare them with experimental data; second, we investigate the temporal evolution of key plasma parameters, identifying transient and steady phases; third, we analyze turbulence characteristics including $E \times B$ shear flow formation and turbulent transport.

\subsubsection{Quasi-steady state profiles}
\begin{figure}
    \centering

    \includegraphics[width=0.49\linewidth]{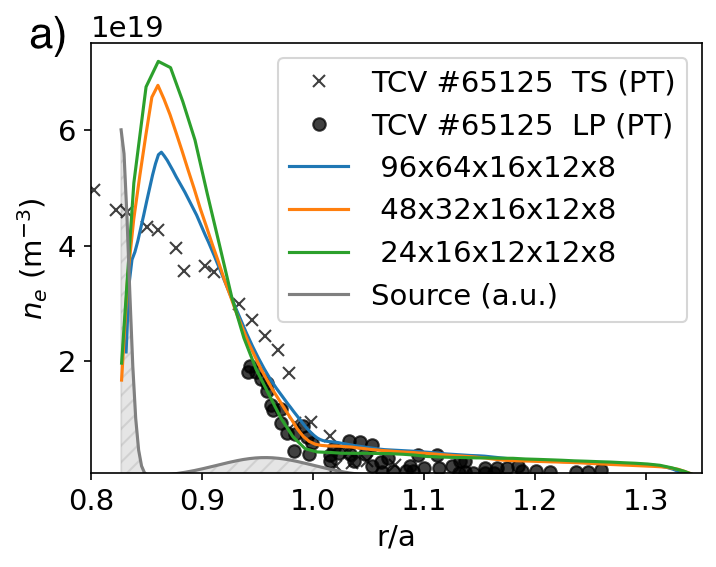} 
    \includegraphics[width=0.49\linewidth]{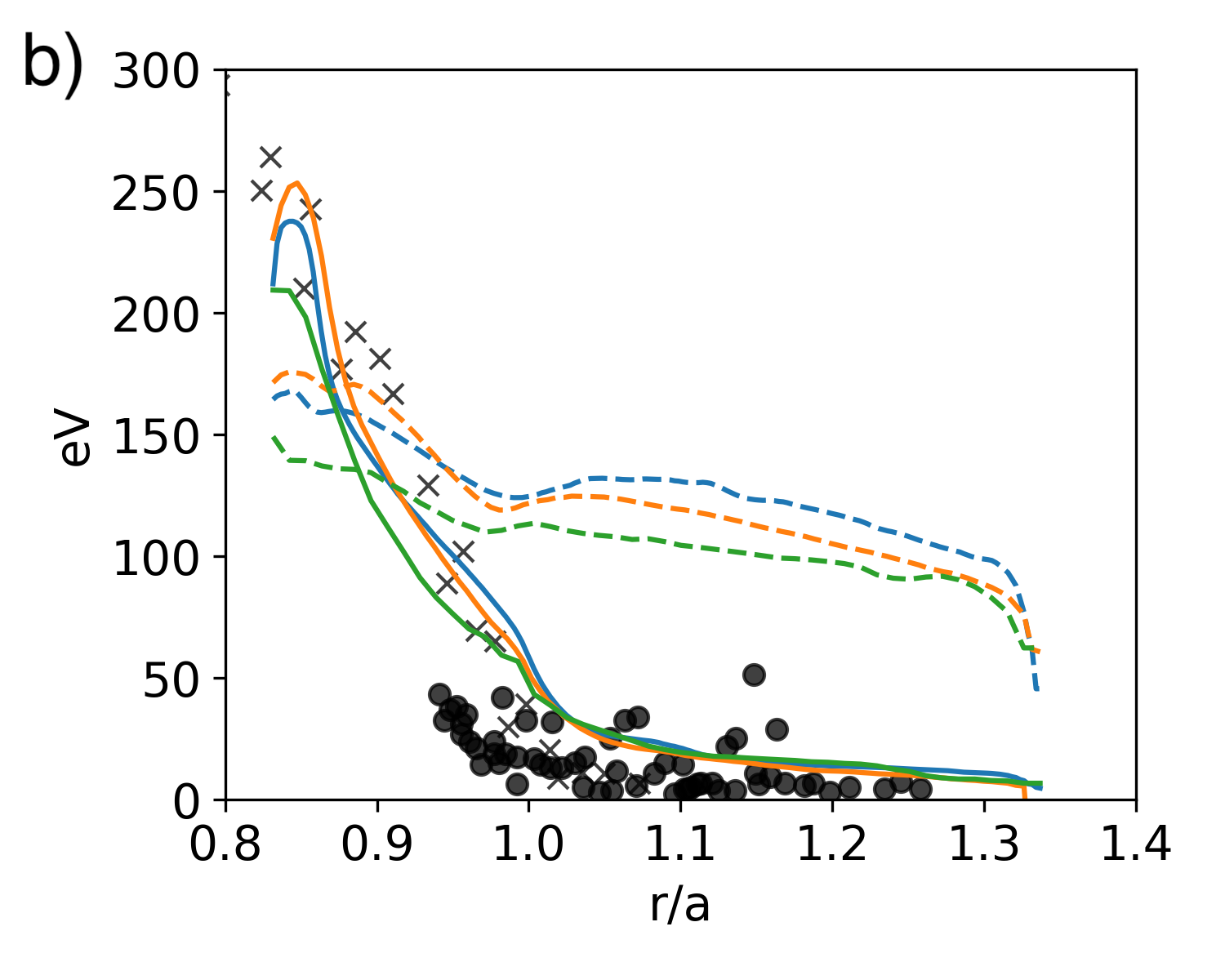}
    \caption{Comparison between the Thomson scattering (TS, $\times$) and Langmuir probe (LP, $\bullet$) electron measurements of TCV \#65125 discharge (PT) with the steady-state density (a), and temperature (b) for electrons (solid) and ions (dashed) obtained with \gkyl{} for different resolutions. The \gkyl{} profiles are averaged over the last 200~$\mu$s of the simulation and taken at the OMP ($z/\pi = 0$), averaged over the binormal direction. We also display the total source profile in panel (a) where the core source is taken at OMP and the recycling source at the limiter position for comparison purposes.}
    \label{fig:PT_kinetic_profiles_vs_exp}
\end{figure}

Figure~\ref{fig:PT_kinetic_profiles_vs_exp} compares \gkyl{} steady-state electron kinetic profiles with TCV \#65125 discharge experimental data (PT), provided by Thomson scattering \citep{blanchard2019thomson} and Langmuir probe diagnostics \citep{boedo2009fastlangmuirprobe,tsui2018langmuirprobe}.
The simulated profiles are taken at the OMP position ($y/\rho_{s0} = 0$, $z/\pi = 0$) and averaged over the last 200~$\mu$s of the simulation.
We observe reasonable agreement between simulations and experiment for each resolution, particularly in the edge and SOL regions.
In the core, the density profile overshoots the experimental data close to the inner radial boundary mostly due to the core source.
The density drops at the inner radial limit due to the absorbing boundary condition.
Closer to the separatrix and in the SOL, the density profile matches the experimental transition from closed to open field line regions within the reported variability.
The electron temperature profile shows somewhat greater resolution sensitivity in the transition region, where higher resolution yields a higher temperature at the LCFS.

We also presents the ion temperature profile, which is not measured experimentally here, in Figure~\ref{fig:PT_kinetic_profiles_vs_exp} (right).
A local minimum is observed just before the last closed flux surface (LCFS). This is due to the propagation of hot eddies along magnetic field lines from the top and bottom of the device into the SOL, locally increasing the ion temperature in these regions.
In the SOL, the ion temperature is significantly higher than the electron temperature due to the parallel streaming timescale being much longer than the radial drifts. 
\modifi{
    We note that the recycling source used in this study does not capture neutral cooling effects, which may lead to an overestimation of the ion temperature in the SOL (see Section~\ref{sec:conclusions} for further discussion of this limitation).
}

The comparison between resolutions indicates that the coarse simulation captures the main profile features, while finer resolutions modestly improve agreement with experimental density and increase edge temperature gradients.
In particular, the higher resolution simulation is able to maintain steeper temperature gradients for both ions and electrons.
This suggests a role for $k_\perp^\text{max}\rho_{s0}\gtrsim 0.6$-scale turbulence: the maximum resolved perpendicular wavenumber is $k_\perp^\text{max}\rho_{s0}\simeq 0.7$ (coarse) versus $\simeq 1.3$ (baseline). Since ITG and TEM instabilities typically develop for $k_\perp\rho_{s0}\gtrsim 0.5$, the higher resolution run can distribute free energy to smaller-scale modes that may transport less efficiently owing to increased sensitivity to $E \times B$ shear.

\modifi{
    We now present a preliminary quantitative assessment of the agreement between simulated and experimental profiles. We compute the coefficient of determination, $R^2$, 
    \begin{equation}
        R^2 = 1 - \frac{\sum_i (y_i - f_i)^2}{\sum_i (y_i - \bar{y})^2},
    \end{equation}
    where $y_i$ are the experimental measurements, $\bar{y}$ their average, and $f_i$ are the simulation values interpolated at the radial locations of the experimental data.
    Table~\ref{tab:r2_comparison} summarizes the $R^2$ values for the electron density, temperature, and pressure profiles in the radial range $r/a > 0.85$ to avoid inner boundary effects.
    The results demonstrate satisfactory agreement between simulation and experiment, with $R^2$ values ranging from 0.87 to 0.96 at the highest resolution and maintaining values above 0.74 even at the baseline resolution.
    Notably, the electron temperature exhibits consistent $R^2$ values across all resolutions ($\sim 0.87$), suggesting that even the coarse resolution captures the dominant temperature profile features.
    The main source of discrepancy appears in the density profile, due to the overshoot near the inner boundary in the simulations.
    The electron pressure profile, obtained from the product of interpolated experimental density and temperature measurements, shows the best agreement across all resolutions, suggesting that disagreements between the two diagnostics may partially explain the individual discrepancies in density and temperature.
    It is worth noting that this metric provides an initial quantification of the agreement and cannot be considered a rigorous validation, as it does not account for experimental error bars.
    A more comprehensive validation study, including fluctuation measurements and detailed uncertainty quantification, will be the subject of future work.
    \begin{table}
    \modifi{
    \centering
    \begin{tabular}{lccc}
        Resolution & $n_e$ & $T_e$ & $p_e$ \\
        $(N_x, N_y, N_z, N_{v\parallel}, N_\mu)$ & $R^2$ & $R^2$ & $R^2$ \\
        $(96,64,16,12,8)$ & 0.88 & 0.87 & 0.96 \\
        $(48,32,16,12,8)$ & 0.74 & 0.86 & 0.82 \\
        $(24,16,12,12,8)$ & 0.55 & 0.87 & 0.75 \\
    \end{tabular}
    \caption{Coefficient of determination ($R^2$) for electron density, temperature, and pressure profiles at different resolutions for the positive triangularity (PT) discharge, evaluated within $r/a \in (0.85, 1.35)$.}
    \label{tab:r2_comparison}
    }
\end{table}

}

\subsubsection{Temporal evolution of plasma parameters}
\begin{figure}
\centering
\includegraphics[width=0.9\textwidth]{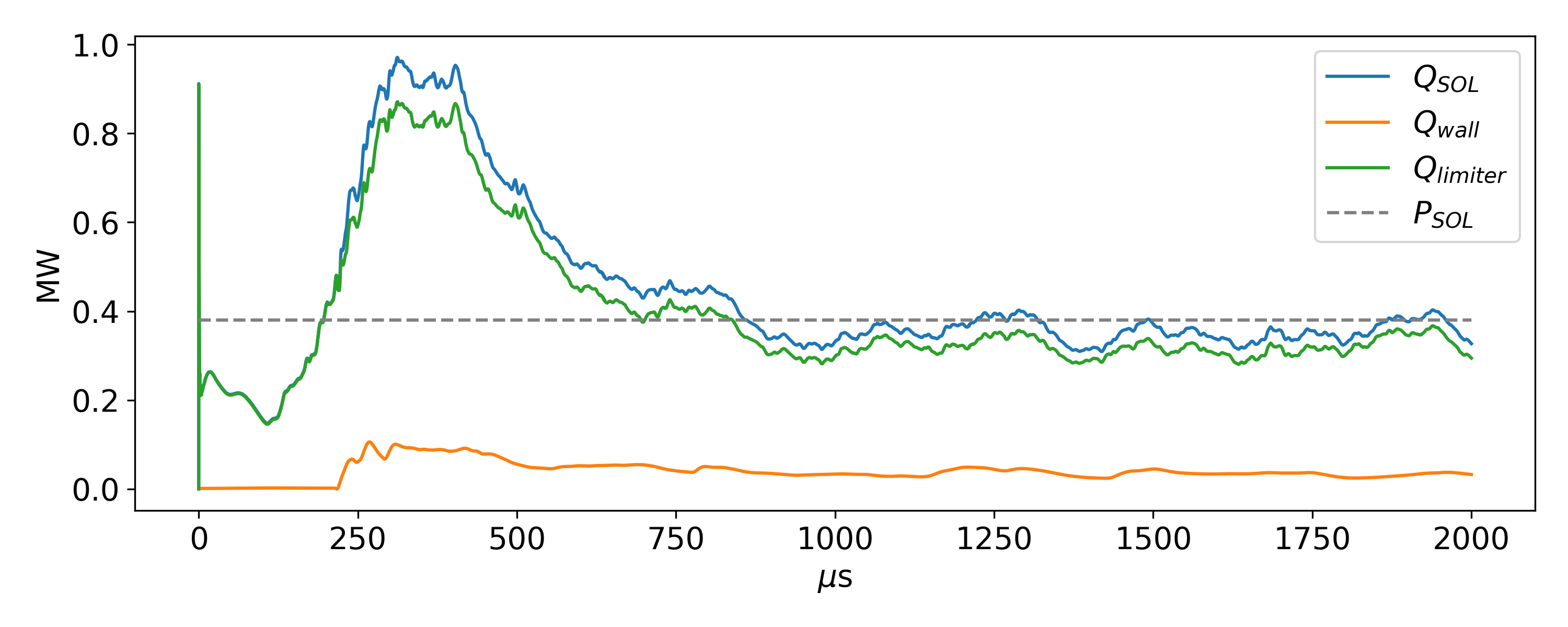} 
\caption{Evolution of the total heat flux through the SOL in \gkyl{} PT simulation (blue), and contributions from the wall (orange) and the limiter (green). The experimental SOL power is shown in gray.
\gkyl{} heat fluxes are here scaled by the inverse of the volume fraction of the flux tube to compare with the experimental input power (see Sec.~\ref{sec:source_setup} for details).}
\label{fig:PT_baseline_energy_out_balance}
\end{figure}

The evolution of the heat flux through the vessel (Fig.~\ref{fig:PT_baseline_energy_out_balance}) illustrates the relaxation timescales.
At the beginning of the simulation, the limiter contribution to the heat flux dominates and exhibits a very short burst of transport ($t\lesssim 5\mu$s) due to the loss of electrons during the formation of the sheath potential.
After this short initial transient, the heat flux increases as turbulence develops into the SOL. 
When the turbulence reaches the wall, corresponding to the non-zero wall heat flux contribution at $t\sim 250\mu$s, the heat losses saturate momentarily.
Feedback from the SOL toward the core is observed after $t\sim 500\mu$s, as the heat flux at the limiter starts to decrease.
Finally, the total heat flux converges toward the target injection power for $t>1$ ms. The ion heat flux is dominant ($\sim 0.15$ MW: $\sim 0.13$ MW limiter, $\sim 0.02$ MW wall). The electron heat flux is lower ($\sim 0.05$ MW at the limiter, negligible at the wall).

\begin{figure}
\centering
\includegraphics[width=0.495\textwidth]{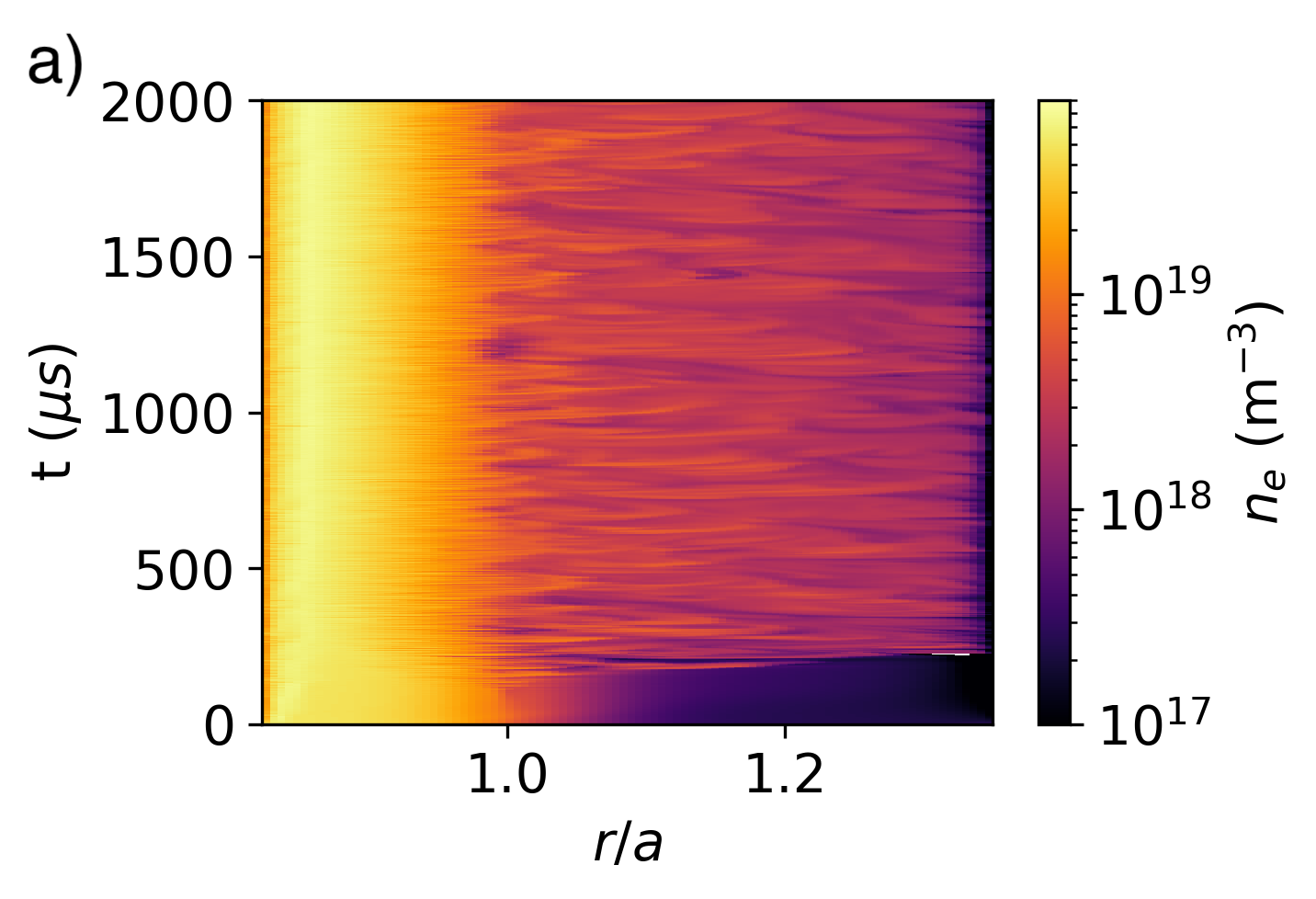}
\includegraphics[width=0.485\textwidth]{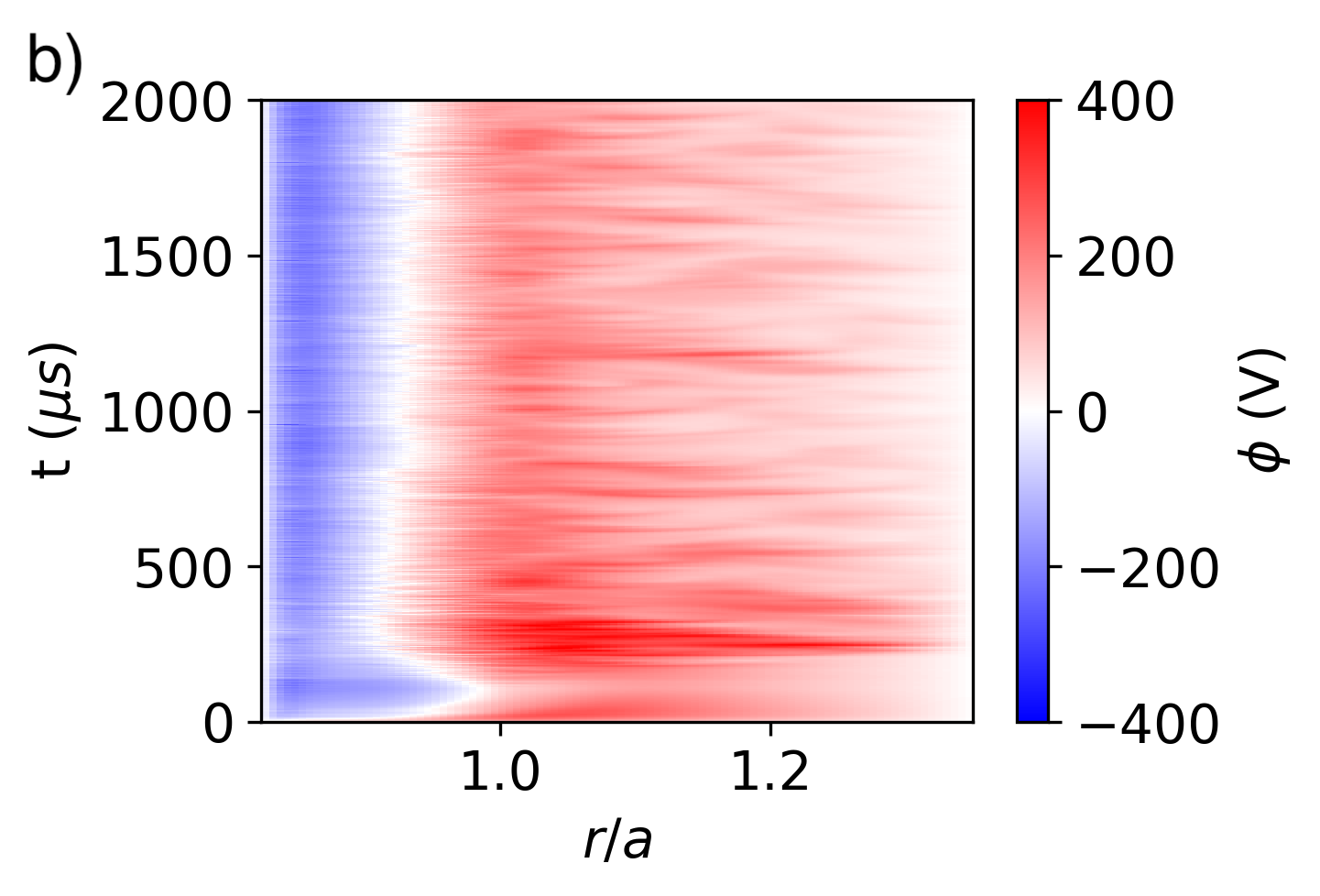}
\includegraphics[width=0.49\textwidth]{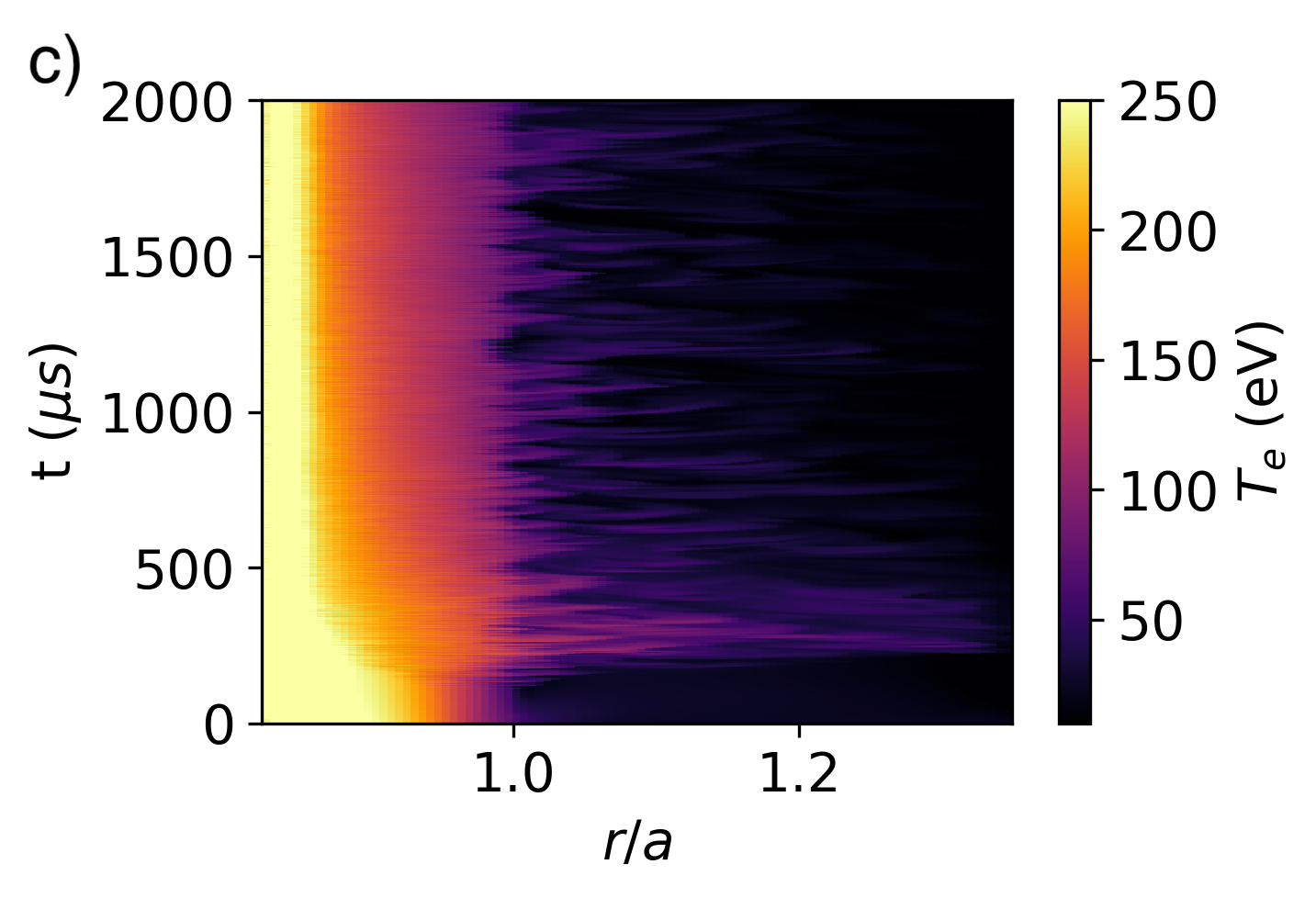}
\includegraphics[width=0.49\textwidth]{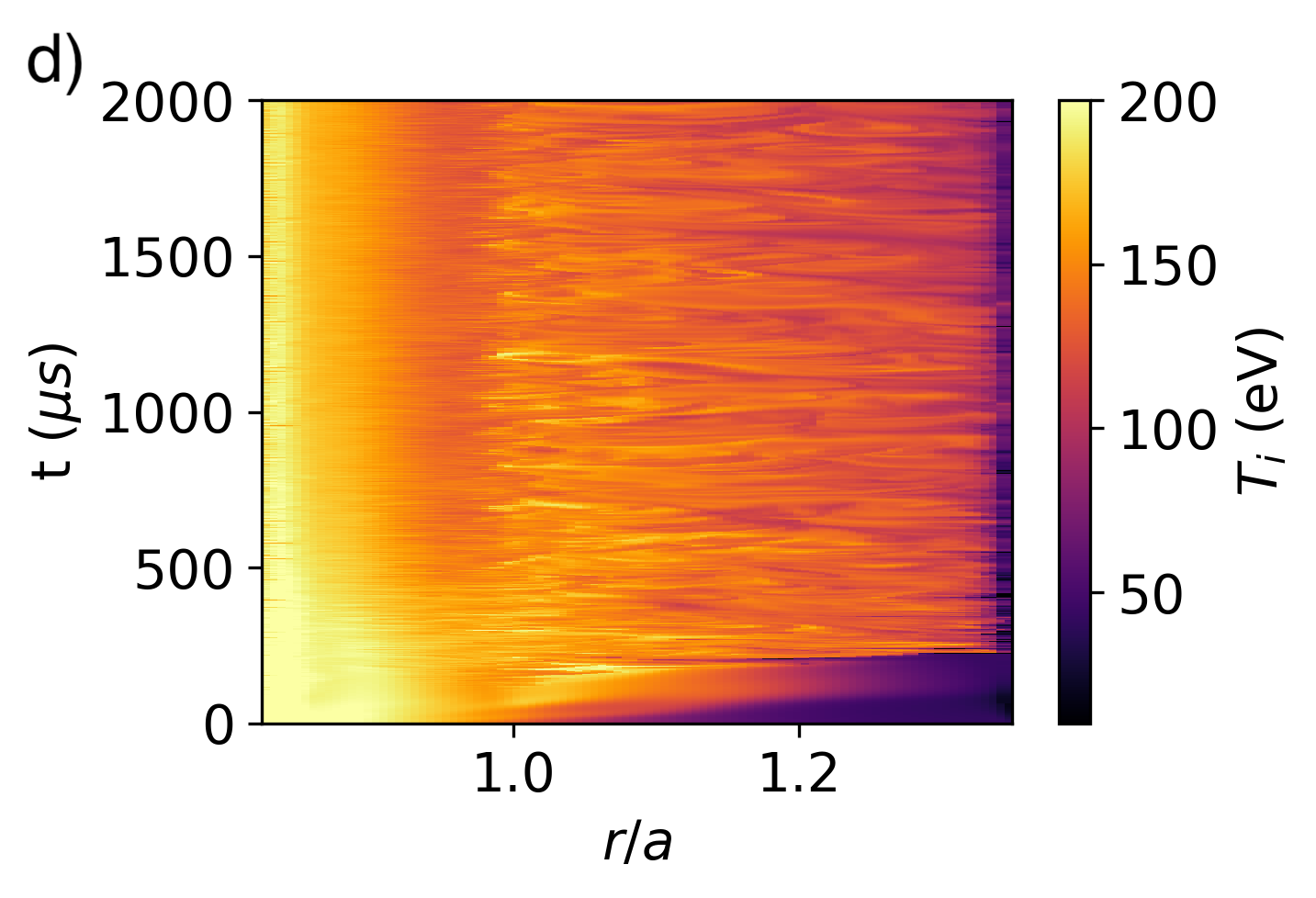}
\caption{Space-time evolution of radial profiles cuts at the OMP ($z=0$) and $y=0$: electron density (a), electrostatic potential (b), electron temperature (c), and ion temperature (d).}
\label{fig:space_time}
\end{figure}

Figure~\ref{fig:space_time} presents the evolution of the plasma kinetic profiles, showing the radial profiles of electron density, electrostatic potential, electron temperature, and ion temperature as functions of time. 
The data is taken at the OMP ($z/\pi = 0$) and constant binormal coordinate $y/\rho_{0i} = 0$.
The electron density evolution (Figure~\ref{fig:space_time}a) shows an inverted gradient close to the inner radial boundary, due to the presence of the absorbing boundary condition.
The electrostatic potential (Figure~\ref{fig:space_time}b) builds up a radial electric field in the core region in response to the polarity of the neoclassical particle flux through the separatrix.
This radial electric field shifts inward when turbulence starts to develop ($t\sim 200\mu$s), correlating with poloidal flow formation and $E \times B$ shear.
This shear layer is sustained in the quasi-steady state, with a radial width much larger than the typical fluctuation scale, suggesting that it is not generated by turbulence but rather by neoclassical effects \citep{Kagan2010neoclassicalpedestal,Kobayashi2015}.
Smaller scale flows are not observed and may be suppressed by the high collisionality regime of this system \citep{Hoffmann2023GyrokineticOperators}.
In the SOL, the potential is mostly constant along the field lines and take a positive value set by the sheath at the limiter boundary.

Electron temperature (Figure~\ref{fig:space_time}c) features a steep gradient near the LCFS due to the fast electron parallel transport along the open field lines to the limiter.
This feature is not present in the ion temperature profile (Figure~\ref{fig:space_time}d), where the parallel streaming timescale is of comparable order to the radial drifts.
Hence, the electron temperature in the SOL is significantly lower.
The relaxation timescale can be estimated from the evolution of the ion temperature profile: turbulence starts to propagate radially outward after roughly 200 $\mu$s, inducing feedback from the SOL to the core. This is consistent with the limiter heat flux saturation after $\sim 250$ $\mu$s (Fig.~\ref{fig:PT_baseline_energy_out_balance}).

\subsubsection{Turbulence dynamics}
\begin{figure}
\centering
\includegraphics[width=0.7\textwidth]{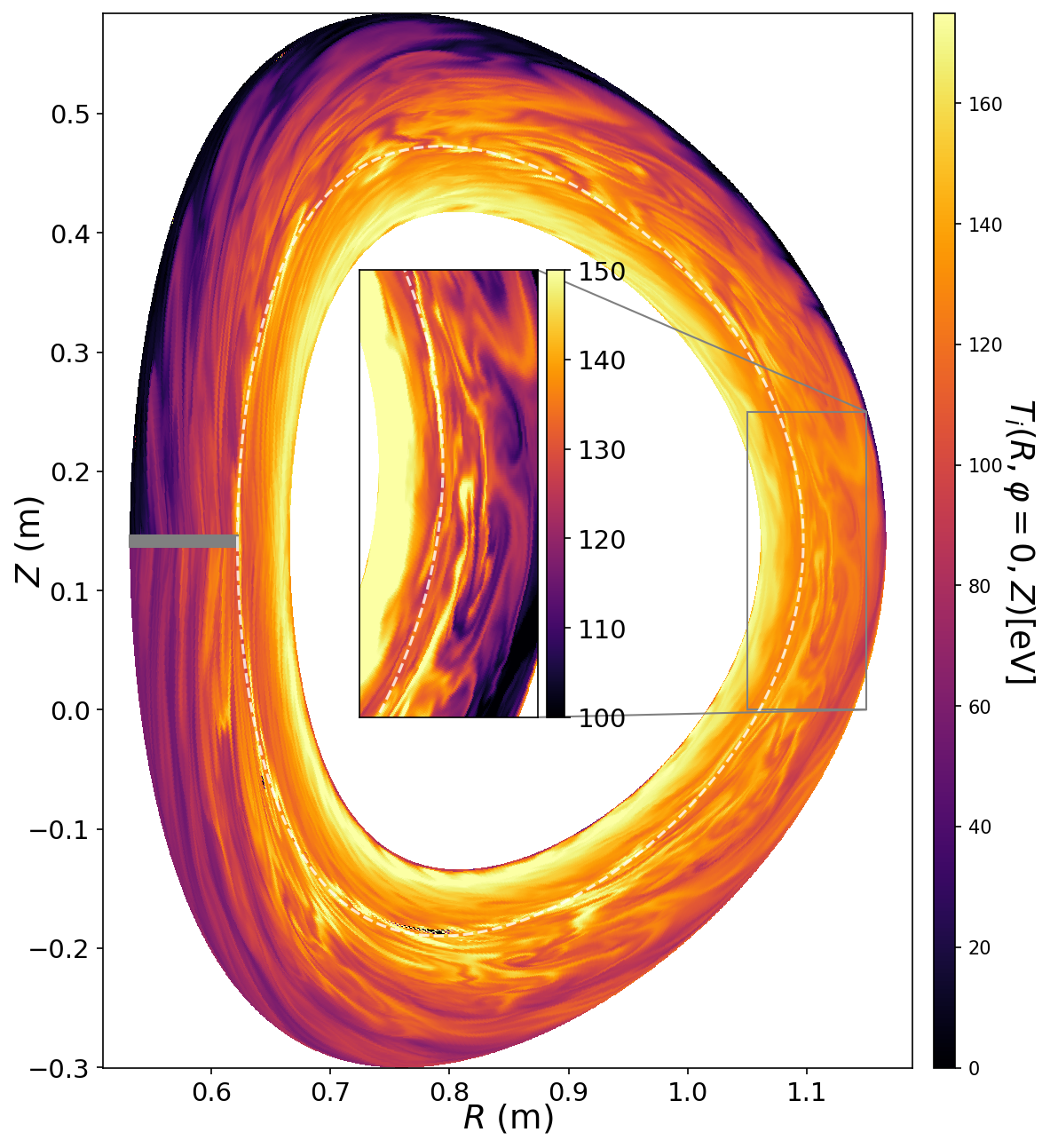} 
\caption{Poloidal cross-section of ion temperature in the PT configuration at an instant in time in the quasi-steady regime ($t\gtrsim 1$~ms). 
The LCFS is indicated by the white dashed line. 
The limiter is represented by a gray rectangle at the high field side.
An inset zooms in on the OMP region with a color map made to strengthen the contrasts.
This plot is obtained by projecting the flux tube domain of \gkyl{} on the poloidal plane at a fixed toroidal angle.}
\label{fig:poloidal_cut}
\end{figure}

Figure~\ref{fig:poloidal_cut} shows a poloidal cross-section of the ion temperature during the quasi-steady state.
The core region hosts the steepest temperature gradient, with a range of approximately 50 eV over a few centimeters \modifi{($R_0/L_{Ti}\sim 10^1$)}.
We report comparable normalized gradients for the ion density and electron temperature profiles which represents a source of free energy for instabilities to develop.
Although this gradient represents a reservoir of free energy for instabilities like ITG and TEM, turbulence in the core region is moderated by $E \times B$ shear associated with the radial electric field (Fig.~\ref{fig:space_time}b).
Fluctuations that survive to the shear, propagate radially outward, eventually reaching the LCFS where they are expelled into the SOL.
This process seeds large-scale structures in the SOL ("blobs").
The blobs are characterized by a higher ion temperature and density compared to the background plasma and represent a major mechanism of particle and energy transport across the SOL by drifting radially outward.
They are also subject to magnetic gradient drifts, which cause preferential motion toward the bottom of the device.
This explains the asymmetry between the top and bottom of the limiter in Figure~\ref{fig:poloidal_cut}.
An estimate from the slope of turbulent filaments in Fig.~\ref{fig:space_time} yields a typical radial blob velocity of order $10^3$ m/s, consistent with scaling arguments \citep{Krasheninnikov2008}.

The inset of Figure~\ref{fig:poloidal_cut} shows a zoomed-in view of the OMP region.
At this location, a localized region of reduced fluctuation amplitude appears just inside the LCFS, with a strong ion temperature gradient, which can be interpreted as a transport barrier.
Outside the LCFS, the SOL displays strong turbulence features exhibiting high temperature blobs.
These blobs appear to originate near the top and bottom of the closed field line region and then travel toward the OMP, contributing to the gap between core and SOL turbulence.

\subsection{Comparison of PT and NT configurations}
\label{sec:results_PT_vs_NT} 
\modifi{
    We now assess the ability of our predictive framework to capture the differences between negative and positive triangularity configurations and investigate the underlying physical mechanisms.

    We focus on TCV discharges \#65125 (PT) and \#65130 (NT), which represent well-diagnosed, comparable L-mode plasmas that differ primarily in triangularity while maintaining similar heating power and plasma current, enabling cleaner isolation of geometric effects on turbulence and transport.
    While the moderate to high collisionality of these L-mode plasmas places them near the fluid-valid regime, several kinetic effects may remain important: self-consistent evolution of the parallel velocity distribution and associated parallel heat transport, kinetic treatment of parallel electron dynamics affecting radial electric field formation, and anisotropic temperature evolution in the SOL where magnetic mirror effects influence particle and energy transport.
    Beyond validating our predictive framework, these simulations establish a baseline for future extension to lower-collisionality regimes where kinetic effects become increasingly dominant \citep{Merlo2023OnTokamaks,Balestri2024PhysicalPlasmas}.
}

\begin{figure}
\centering
\includegraphics[width=0.49\textwidth]{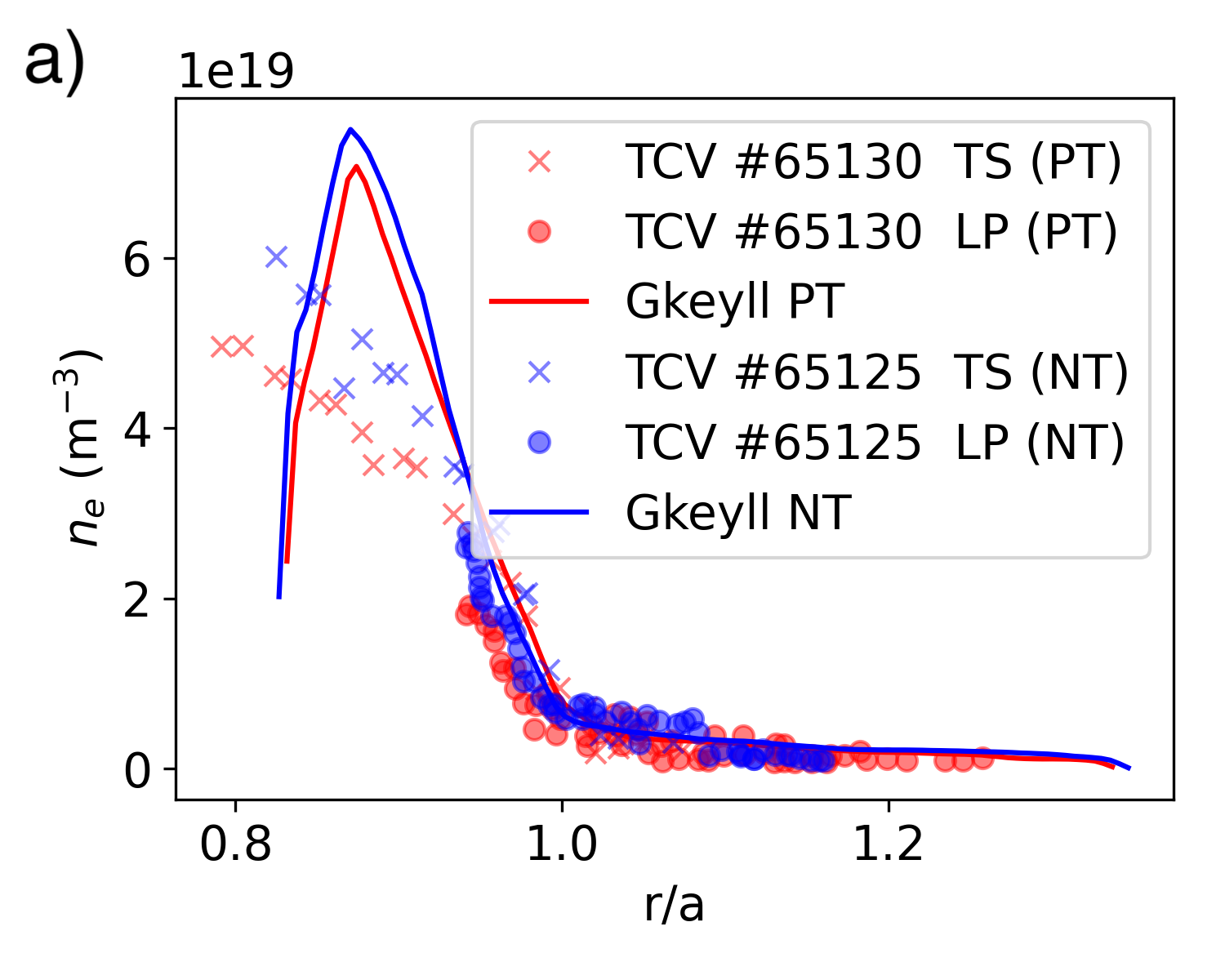} 
\includegraphics[width=0.49\textwidth]{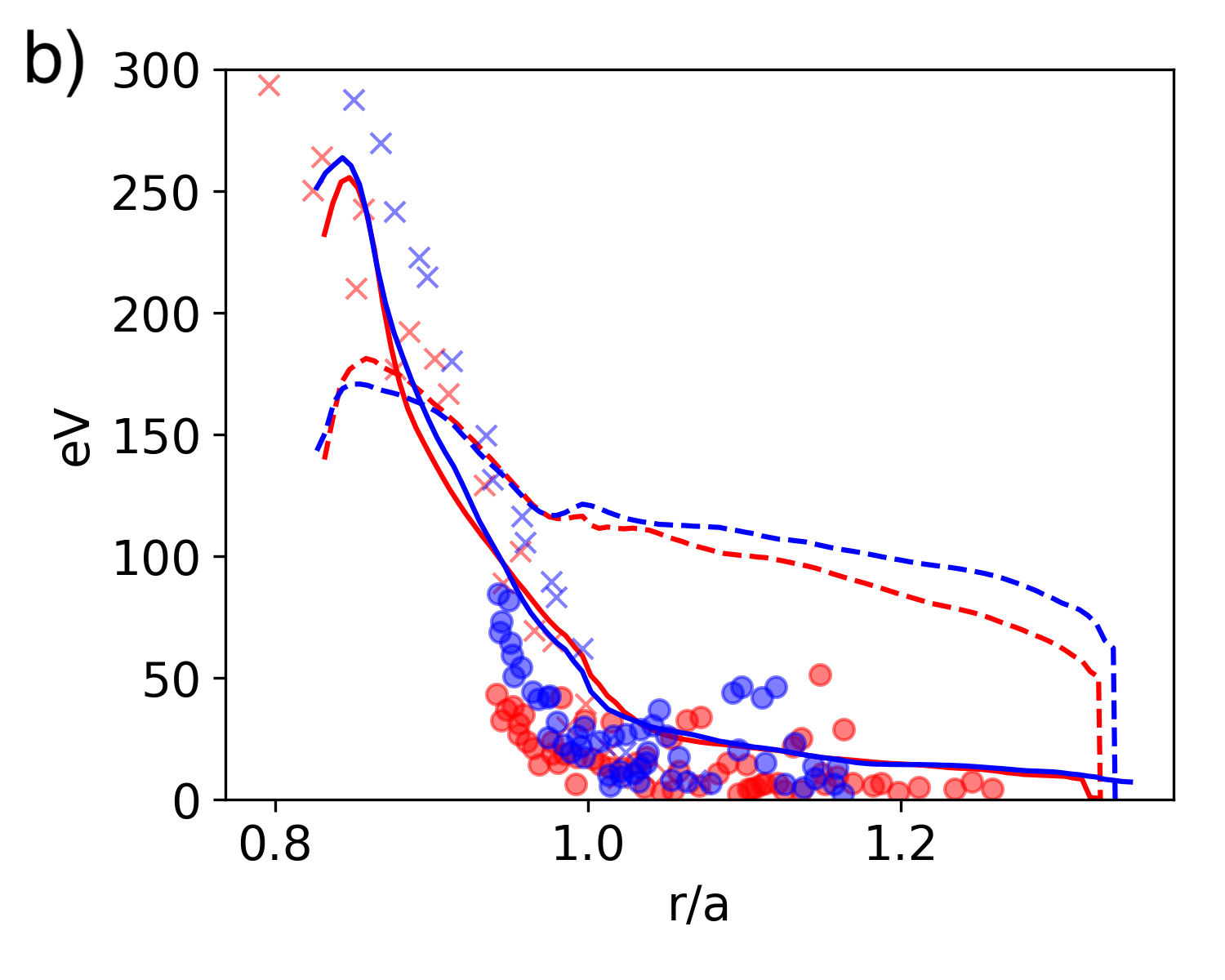} 
\caption{Comparison of the electron density (a) and temperature profiles (b) for PT (red) and NT (blue) configurations averaged over 200$\mu$s in the quasi-steady state of the baseline resolution simulation at the OMP ($z=0$) and averaged along $y$.
The electron temperature is shown in solid lines, while the ion temperature is shown in dashed lines.
The resulting profiles are compared to experimental electron data obtained from Thomson scattering ($\times$) and Langmuir probe ($\bullet$) diagnostics.}
\label{fig:profiles_PT_vs_NT}
\end{figure}

Figure~\ref{fig:profiles_PT_vs_NT} compares the experimental measurements of electron density and temperature profiles with the quasi-steady state profiles at OMP obtained with \gkyl{} for both configurations, using the baseline resolution.
The experimental measurements show that the NT configuration exhibits an increase in electron density for normalized minor radius $r/a \lesssim 0.9$, while SOL profile data are less conclusive. The electron temperature profiles show weaker variation between NT and PT; at smaller radii, the NT configuration exhibits slightly higher electron temperatures compared with PT.
The \gkyl{} simulations capture several experimental trends, with some quantitative differences. At the OMP, the simulations reproduce subtle NT–PT differences. The electron density profiles are consistent with improved core confinement in the NT case. The simulated electron temperature profiles are similar between configurations, while the ion temperature shows a relative increase in the NT SOL region, suggesting altered ion energy transport there.

\begin{figure}
    \centering
    \includegraphics[width=0.49\linewidth]{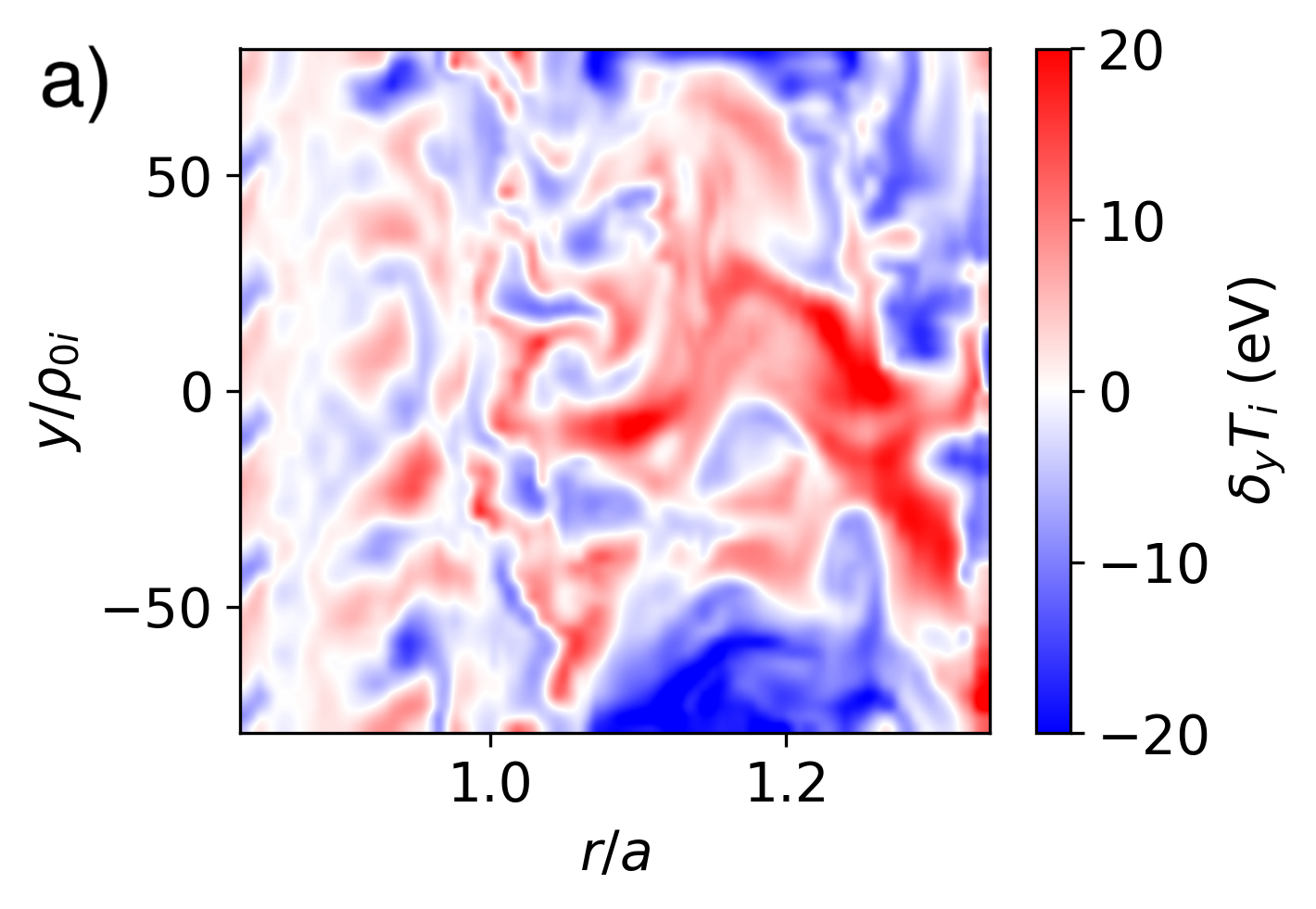} 
    \includegraphics[width=0.49\linewidth]{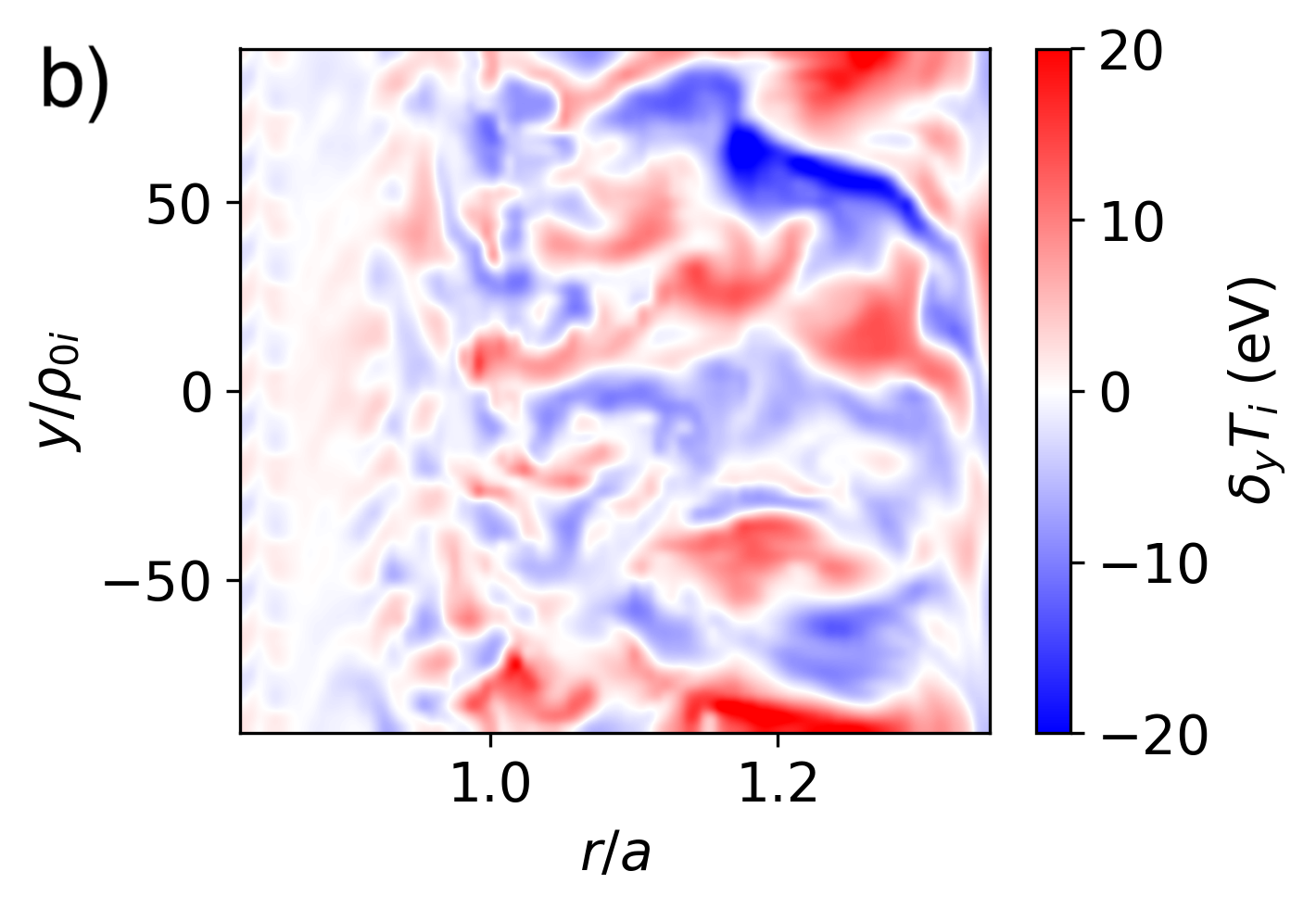} 
    \caption{Ion temperature fluctuation at the OMP ($z=0$) for PT (a) and NT (b) configurations at the baseline resolution for $t=1500~\mu$s.
    The fluctuation is defined as $\delta_y T_i = T_i - \langle T_i \rangle_y$, where $\langle \cdot \rangle_y$ denotes an average in the binormal direction.}
    \label{fig:dTi_NT_vs_PT}
\end{figure}

In the closed field line region, plasma transport is predominantly governed by turbulent $\mathbf{E} \times \mathbf{B}$ drift. Figure~\ref{fig:dTi_NT_vs_PT} illustrates ion temperature fluctuations at the OMP for both configurations. Both PT and NT exhibit a region of reduced fluctuation amplitude; this region is somewhat more pronounced in the NT case.
The PT configuration exhibits larger amplitude fluctuations than NT (Fig.~\ref{fig:dTi_NT_vs_PT}); a similar qualitative trend is seen in electrostatic potential, electron density, and electron temperature fluctuations (not shown). The NT case shows finer-scale structure, which could contribute to reduced transport via gyro-Bohm-like scaling.

\begin{figure}
    \modifi{
    \centering
    \includegraphics[width=0.49\textwidth]{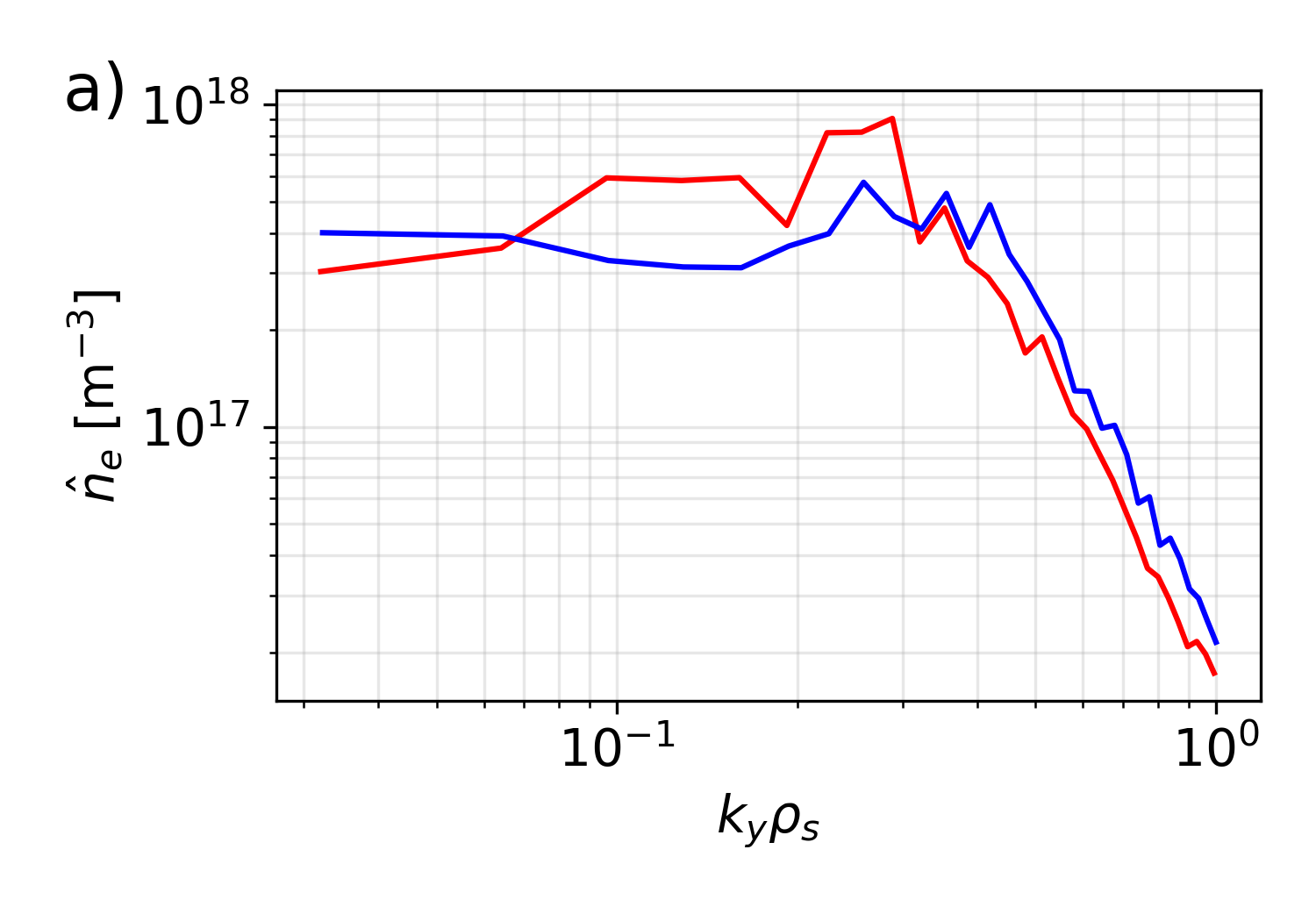}
    \includegraphics[width=0.49\textwidth]{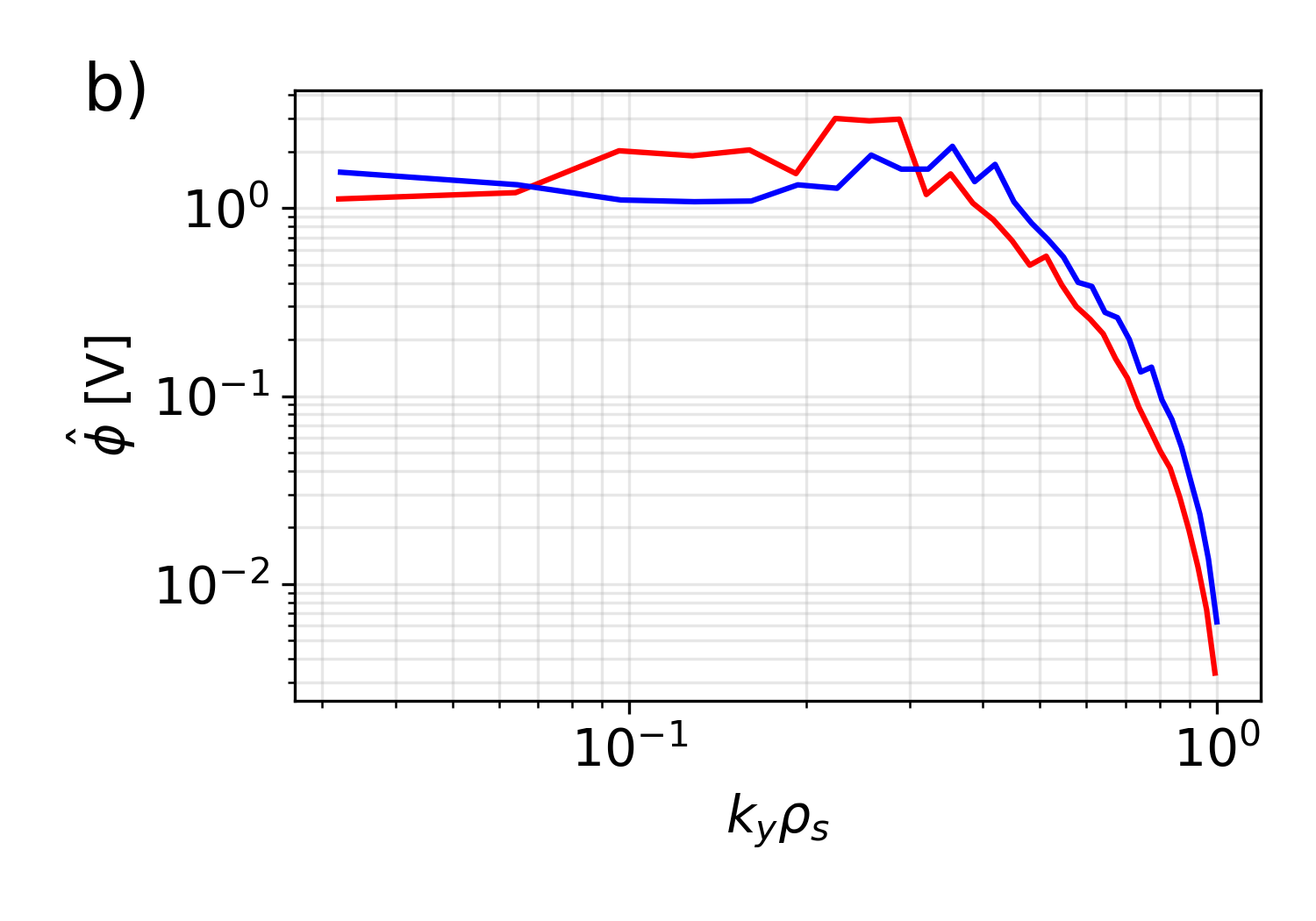}
    \includegraphics[width=0.49\textwidth]{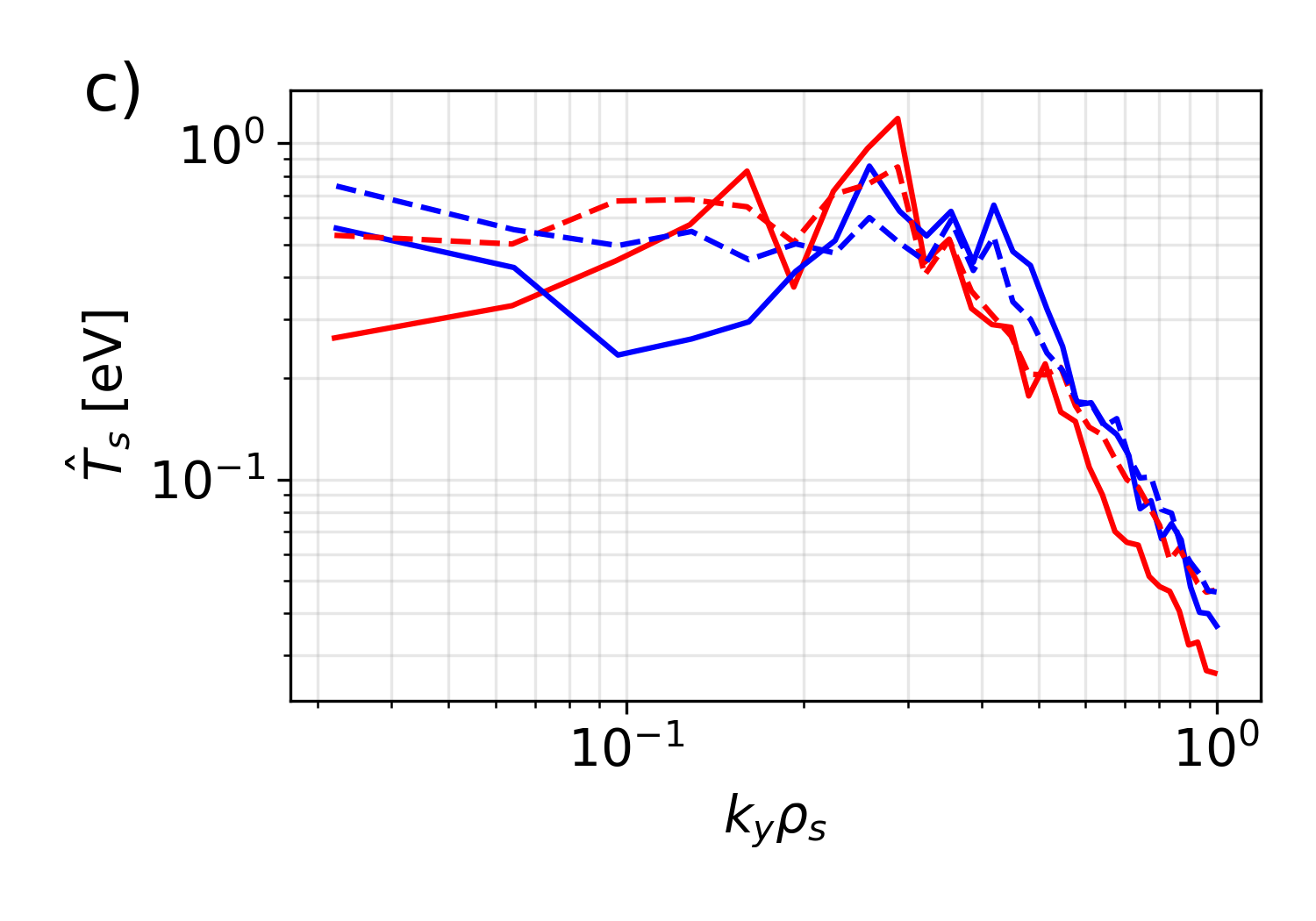}
    \includegraphics[width=0.49\textwidth]{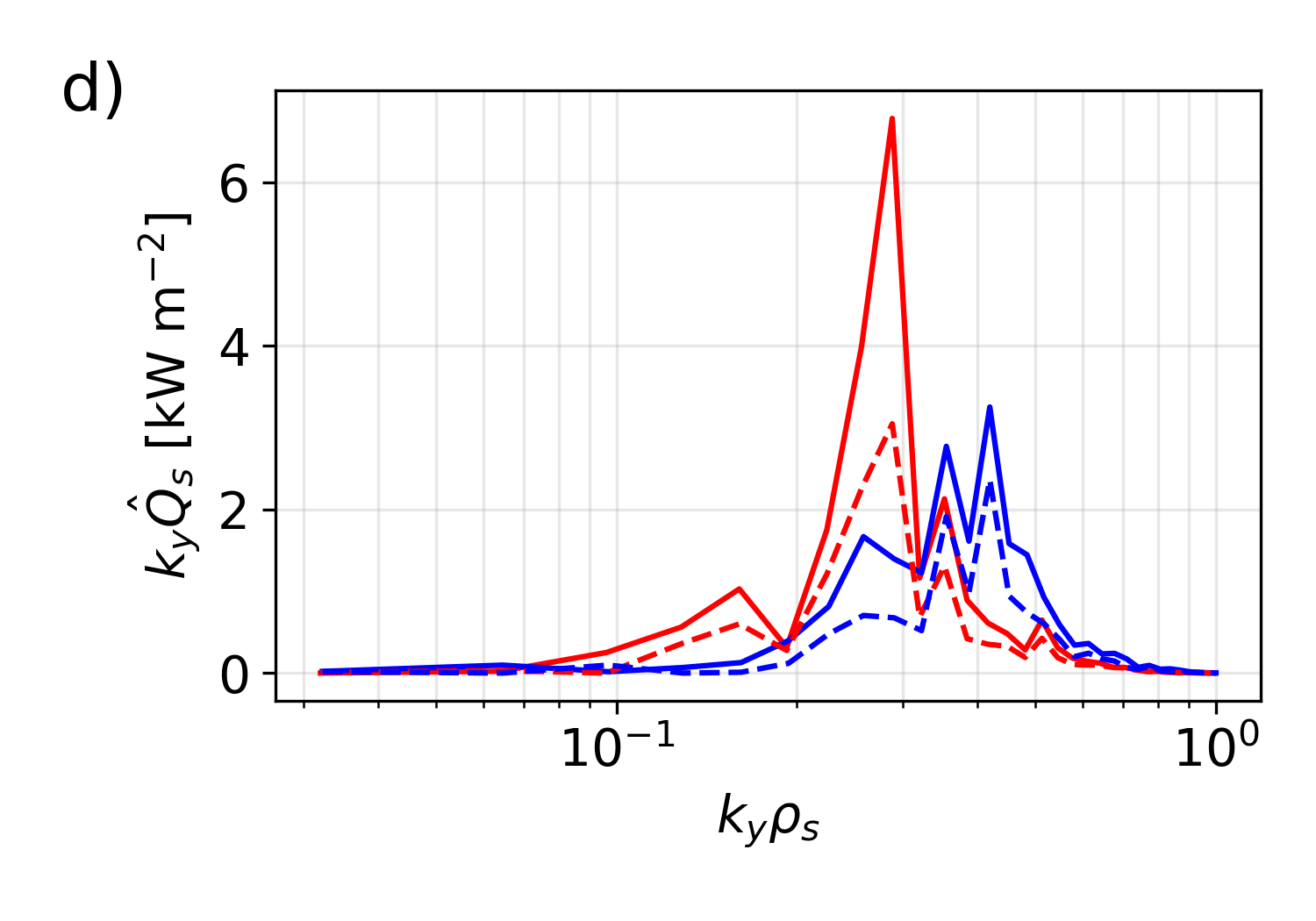}
    \caption{Fluctuation spectra of density (a), electrostatic potential (b), temperature (c), and total heat flux (d) at the OMP ($z=0$) for PT (red) and NT (blue) configurations. The solid lines indicate the electron quantities, while the dashed lines indicate the ion quantities. The spectra are computed at $r/a=0.9$ and averaged over $100~\mu$s in the quasi-steady state of the fine resolution simulations. The spectra are computed using cell averages (the zeroth DG coefficient) to avoid aliasing issues that can arise from the interpolation procedure, which effectively limits the displayed wavenumber range.}
    }
    \label{fig:fluct_spectra}
\end{figure}

\modifi{
    The spectra of fluctuations at the OMP and $r/a=0.9$ are presented in Fig.~\ref{fig:fluct_spectra}.
    We confirm that the PT configuration has higher fluctuation levels in the low-wavenumber range ($k_\perp \rho_s \lesssim 0.5$) for all quantities considered (density, $\hat n_e$, electrostatic potential, $\hat \phi$ and temperatures, $\hat T_e$ and $\hat T_i$). In contrast, the NT configuration exhibits relatively enhanced fluctuations at higher wavenumbers ($k_\perp \rho_s \gtrsim 0.5$), particularly for density and electrostatic potential. This shift toward smaller scales in the NT case may contribute to the observed reduction in overall transport levels, as smaller-scale turbulence is less effective at driving cross-field transport.

    We now turn to the turbulent heat flux spectra,
    \begin{equation}
    \hat{Q}_{\text{tot},s}(x,k_y,t) = \frac{3}{2} \langle T_s \rangle_{0} \, k_y \, \text{Im}\left\langle \frac{\tilde{n}_s^* \, \tilde{\phi}}{B} \right\rangle_{\text{yz}}+\frac{3}{2} \langle n_s \rangle_{0} \, k_y \, \text{Im}\left\langle \frac{\tilde{T}_s^* \, \tilde{\phi}}{B} \right\rangle_{\text{yz}}
    \end{equation}
    where $\langle \cdot \rangle_{yz}$ denotes the instantaneous flux-surface average, $\langle \cdot \rangle_{yzt}$ denotes the time- and flux-surface-average, and $\tilde{g}= \tilde{g}(x,k_y,t)$ represents the fluctuation spectrum of a field $g$.
    Figure \ref{fig:fluct_spectra}d shows that the turbulent heat flux spectrum for the PT configuration peaks at lower wavenumbers compared to the NT case, indicating that larger-scale turbulent structures dominate heat transport in PT.
    We also find that the electron fluctuations and heat flux spectra tend to have higher amplitudes than the ion counterparts, indicating that electron dynamics play a significant role in the turbulence characteristics in these configurations, which is consistent with \cite{ulbl2023tcvx21genex}.
    This also suggests that TEM is the dominant instability drive in these simulations, as also found in \cite{ulbl2023tcvx21genex}. 
    We confirm this hypothesis by performing linear GK simulations with the \gyacomo{} code \citep{Hoffmann2023GyrokineticShift}, using logarithmic gradients obtained using the time- and flux-surface-averaged profiles from the nonlinear \gkyl{} simulations as input. In particular, \gyacomo{} simulations confirm the TEM dominance in both configurations at $r/a=0.9$ and $r/a=0.95$, with only minor differences in linear growth rates between PT and NT. We note that the logarithmic gradients used here are of the order of $R_0/L_{n} = 40$--$80$ and $R_0/L_{Te} = 20$--$30$, which is much higher than the DIII-D edge case studied in \cite{Hoffmann2025InvestigationSimulations}, where ITG and TEM drives were found to compete more closely.
}


\begin{figure}
\centering
\includegraphics[width=0.49\textwidth]{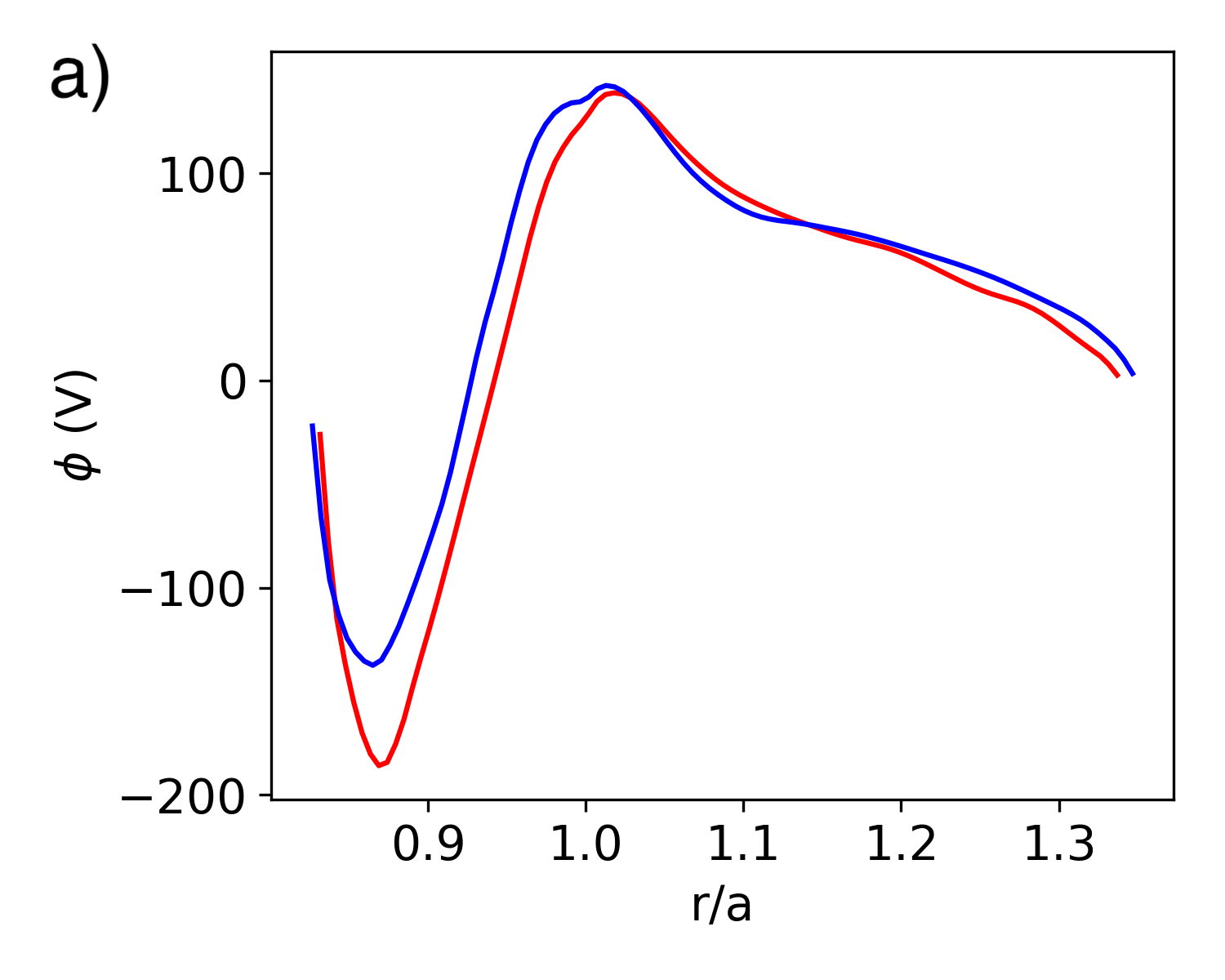} 
\includegraphics[width=0.49\textwidth]{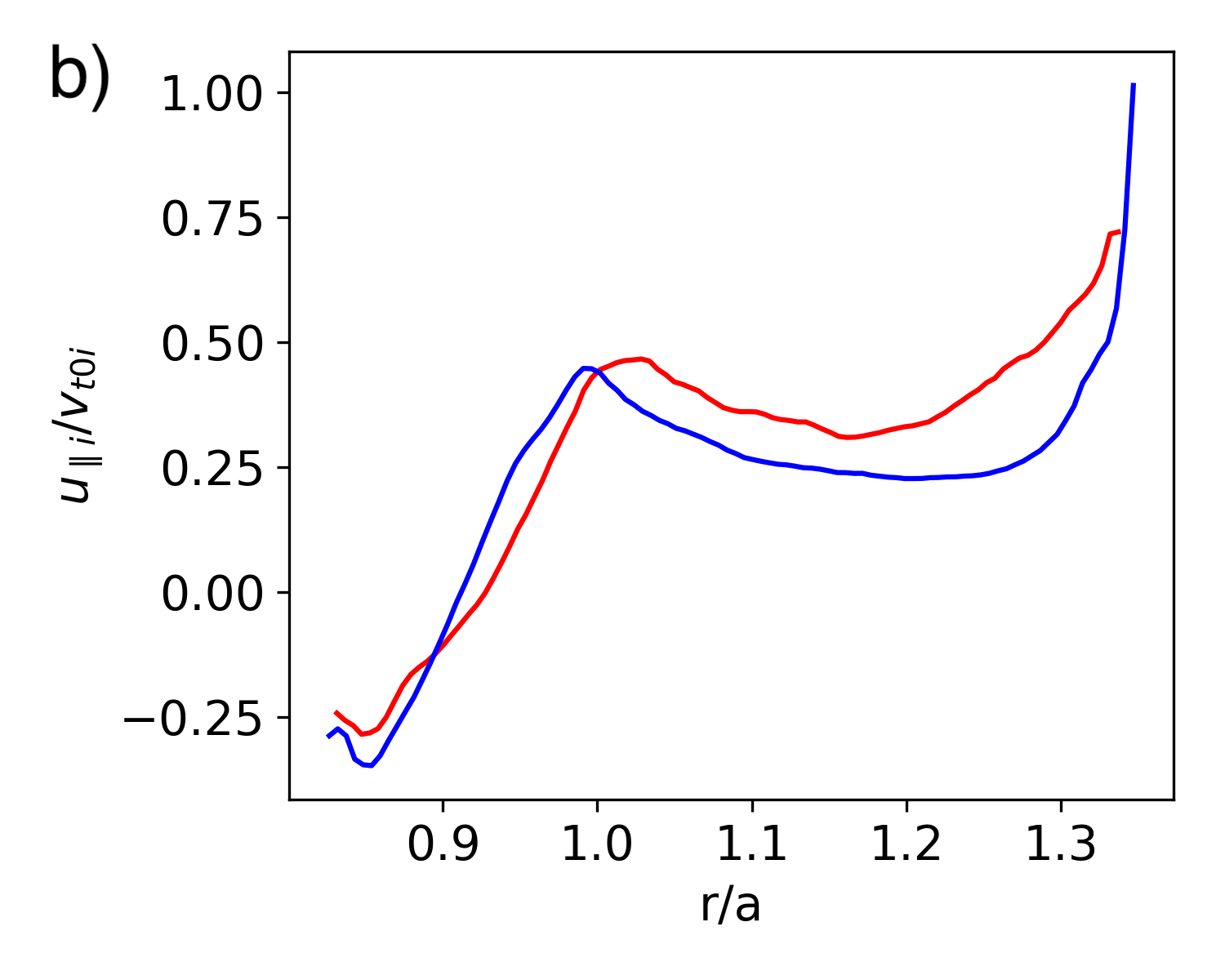} 
\caption{Electrostatic potential (a) and ion parallel velocity (b) for PT (red) and NT (blue) configurations averaged over 200$\mu$s in the quasi-steady state of the baseline resolution simulation at the OMP ($z=0$) and averaged over $y$.}
\label{fig:profiles_PT_vs_NT_ui_phi}
\end{figure}

The full-$f$ formulation allows self-consistent large-scale electric fields, typically inward in the core and outward in the SOL. Time-averaged electrostatic potential profiles at the OMP (Fig.~\ref{fig:profiles_PT_vs_NT_ui_phi}) fitted in the outer core region ($0.85 < r/a < 1.0$) yield estimated radial shearing rates $2.0 \times 10^5~\text{s}^{-1}$ (PT) and $2.4 \times 10^5~\text{s}^{-1}$ (NT), an increase of about 20\%. This enhanced shearing may contribute to turbulence suppression in the NT case.

In the SOL region, transport dynamics differ from the closed field line region, being dominated by parallel streaming to the limiter and magnetic gradient-B drift toward the vessel wall. Geometric differences between configurations can modify connection length (proportional to safety factor $q$), influencing required parallel velocities for comparable cooling. This is reflected in Fig.~\ref{fig:profiles_PT_vs_NT_ui_phi}, where the parallel velocity profiles differ between configurations.
The parallel and perpendicular temperature components provide insight into SOL transport. The ion parallel temperature in the SOL is substantially lower than the perpendicular temperature, consistent with efficient parallel losses selecting higher magnetic moment particles, for which $\mathbf{E} \times \mathbf{B}$ and gradient-B drifts become relatively more important.
Triangularity appears to impact the perpendicular ion temperature, particularly in the far SOL. The NT configuration exhibits higher perpendicular values. Higher perpendicular temperature increases gradient-B drift velocity toward the wall; the observed differences are consistent with altered cross-field transport under NT geometry but further analysis would be required to isolate causality.
Particles with higher magnetic moments experience stronger mirror forces that can inhibit parallel transport to the limiter. 

\modifi{
    The differences in perpendicular temperature and transport characteristics between PT and NT configurations motivate examining how triangularity affects the power exhaust partitioning between wall and limiter boundaries.
    Our power balance analysis reveals a measurable shift: NT directs 8.1\% of total power to the wall and 91.9\% to the limiter, compared to 6.5\% and 93.5\% for PT. This represents a 25\% relative increase in wall power loading for NT. While the absolute change is modest, such redistribution could be relevant for divertor and first-wall design in reactor scenarios. However, broader parameter scans and inclusion of additional physics (electromagnetic effects, neutral dynamics, realistic divertor geometry) would be needed to assess the generality of this trend.
}

\begin{figure}
\centering
\includegraphics[width=0.49\textwidth]{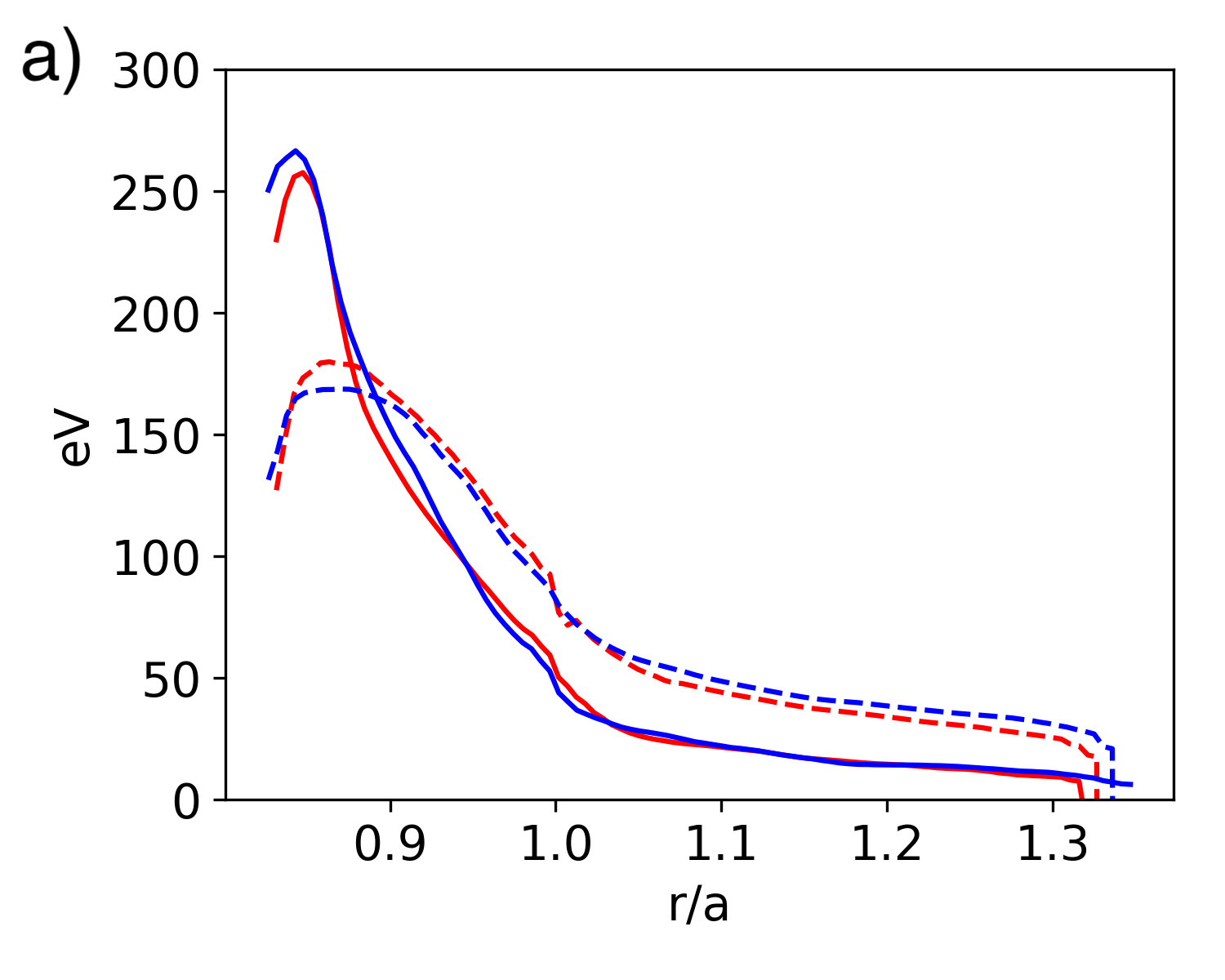} 
\includegraphics[width=0.49\textwidth]{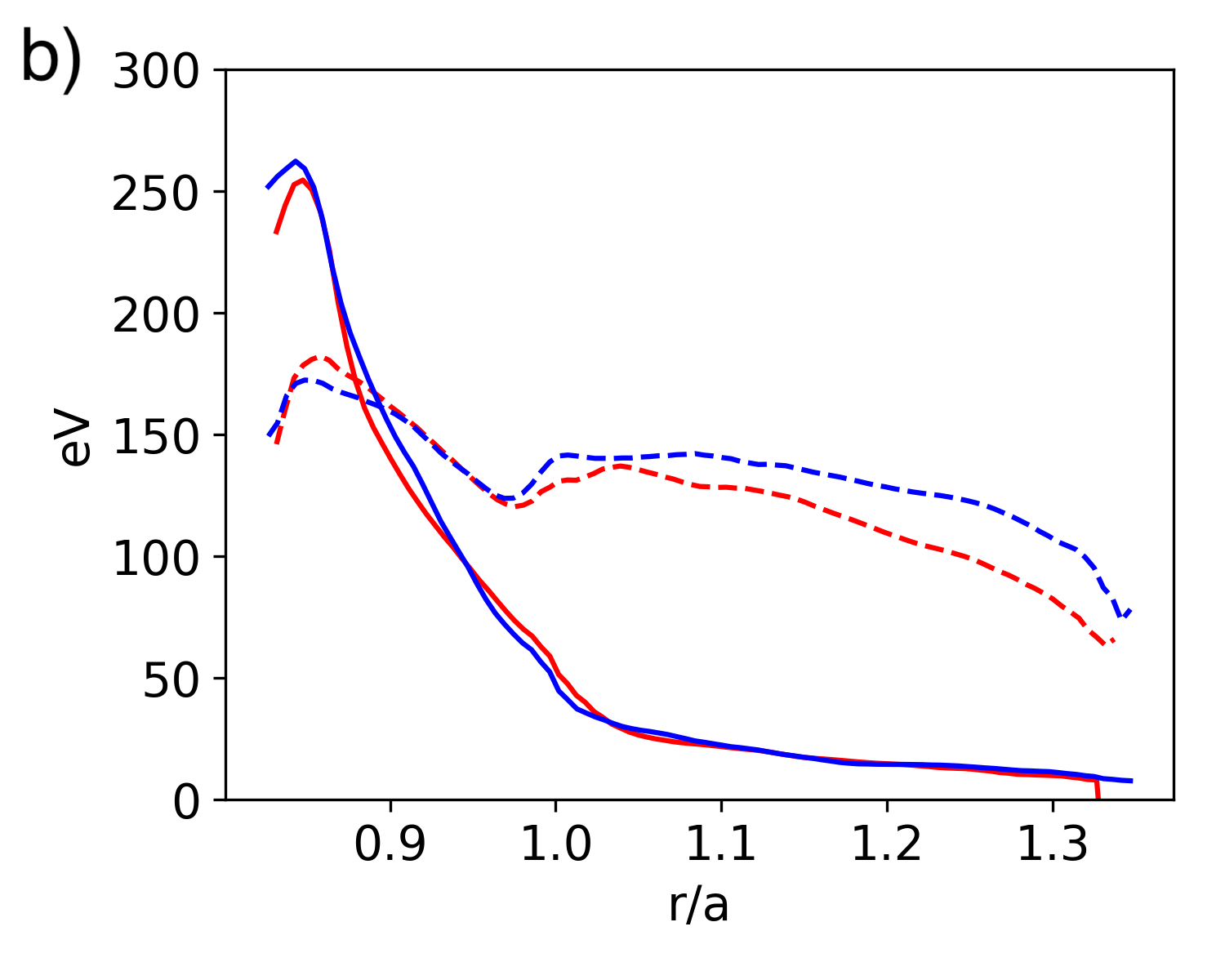} 
\caption{Parallel (a) and perpendicular (b) temperatures for PT (red) and NT (blue) configurations averaged over 200$\mu$s in the quasi-steady state of the baseline resolution simulation at the OMP ($z=0$) and averaged over $y$.
The electron temperature is shown in solid lines, while the ion temperature is shown in dashed lines.}
\label{fig:profiles_PT_vs_NT_Tpar_Tperp}
\end{figure}

\section{Conclusions}
\label{sec:conclusions}

This paper presents full-$f$ gyrokinetic edge and scrape-off layer (SOL) turbulence simulations based solely on three inputs: power, magnetic equilibrium, and total particle number. The framework evolves turbulence without prescribing temperature or density profiles within the simulation domain, providing profile predictions subject to the underlying model assumptions. This departs from previous approaches that impose boundary or profile constraints, though further validation across regimes will be required to assess generality.

This advanced predictive capability relies on the implementation of an adaptive sourcing scheme in the \gkyl{} gyrokinetic code. The scheme emulates power transport from the core to the edge without net particle injection while approximating recycling of particles lost to material surfaces. This zero net particle injection is particularly suitable for Ohmic- and electron cyclotron-heated plasmas, while NBI effects can be captured by this source approach in an even simpler manner through direct particle injection. This approach represents a compromise between computationally intensive kinetic neutral modeling and simpler prescribed-neutral treatments.

Comparison with available measurements for TCV discharge \#65125 shows reasonable agreement for electron density and temperature profiles within the limitations of the diagnostics and simulation assumptions. The simulations reproduce characteristic phenomena such as turbulent blob development in the SOL and formation of a radial electric field. Power accounting indicates consistency between injected and exhausted power under the adaptive sourcing implementation.

We further evaluate the adaptive sourcing approach by comparing positive triangularity (PT) and negative triangularity (NT) discharges (\#65125 and \#65130). The simulations reproduce an NT electron density increase in the core region. They also indicate an approximately 20\% higher $\mathbf{E} \times \mathbf{B}$ shearing rate in the outer edge region, along with a modest change in power partition (wall vs limiter). These observations are consistent with proposed NT confinement mechanisms. 
\modifi{
    The spectrum of density and temperature fluctuations at the outer mid-plane (OMP) also indicate a reduction of turbulent activity at mid-to-high wavenumbers in the NT case, consistent with reduced turbulent transport. 
    The electron contributes dominantly to the turbulent heat flux in both configurations, indicating the prevalence of TEM driven turbulence. This is confirmed by performing linear \gyacomo{} simulations using the logarithmic gradients obtained from the nonlinear \gkyl{} simulations as input.}

\modifi{
    However, several limitations remain in our Gkeyll simulation framework: (i) a simplified recycling model lacking neutral cooling physics, (ii) the omission of radial variation in shaping parameters in the Miller geometry that may influence inner-edge transport, (iii) the lack of higher-order FLR effects, and (iv) electromagnetic fluctuations.
    
    Addressing the recycling model, a more accurate neutral treatment \citep{Bernard2022KineticModelingNeutral} would likely have a larger impact on the ion temperature profile in the SOL, as the current recycling source does not capture neutral cooling effects.
    Regarding geometry, improvements are being made to generate the computational domain directly from EFIT equilibrium reconstructions \citep{Liu2025collision}, which reduces uncertainties associated with the accuracy of our Miller parametrization.
    Additionally, the recent implementation of X-point geometry in \gkyl{} \citep{shukla2025constructingfieldalignedcoordinate} will enable turbulence simulations of diverted configurations in the future.
    This capability opens the path toward validation against the TCV-X21 dataset \citep{Oliveira2022ValidationCase} which is a reference benchmark for edge turbulence codes \citep{michels2021,ulbl2023tcvx21genex,Wang2024solpsitertcvx21,bufferand2024soledge3xtcvx21,dudson2025validationhermes3turbulencesimulations}.
    
    Higher-order FLR and electromagnetic effects will be required to explore regimes with higher pressure gradients, such as in spherical tokamaks and the H-mode pedestal region. 
    Recent work \citep{balestri2025interplaytriangularitymtm} indicates also that micro tearing modes may detrimentally impact NT confinement, motivating future investigation of global full-f electromagnetic gyrokinetic simulations in NT configurations.
    
    Beyond addressing these limitations, we plan to investigate further improvements to the boundary conditions, such as relaxing the $\phi=0$ constraint at the inner boundary, which may reduce the shear flow region suspected to suppress turbulence in that area, and exploring surrogate models for sheath physics at the limiter boundary \citep{Geraldini2024SheathPlasmas}.
    The adaptive source scheme could also be extended to include momentum injection to model the torque from NBI-driven discharges.
    Additionally, a more comprehensive assessment of NT effects requires broader parameter scans in more controlled settings to isolate causality. A direct comparison with GBS NT studies \citep{Riva2020ShapingEffects} would help in this regard, clarifying the role of kinetic effects.
}

\section{Acknowledgements}
\label{sec:acknowledgements}
The authors thank Jimmy Juno for his assistance with code development, the rest of the \gkyl{} team for helpful conversations, the TCV team for providing experimental data and support, and Felix Parra and the PPPL theory group for their valuable discussions.
This work is supported by a DOE Distinguished Scientist award, the CEDA SciDAC project and other PPPL projects via DOE Contract Number DE-AC02-09CH11466 for the Princeton Plasma Physics Laboratory.

\newpage


\bibliographystyle{vancouver}
\bibliography{references}
\end{document}